\documentclass[aps,prb,twocolumn,floatfix,longbibliography,superscriptaddress]{revtex4-2}

\usepackage[T1]{fontenc}
\usepackage{amsmath,amssymb,bm}
\usepackage{graphicx}
\usepackage{grffile}
\usepackage{xcolor}
\usepackage{hyperref}
\hypersetup{hidelinks}
\usepackage{mathtools}
\usepackage{microtype}
\usepackage{placeins}
\usepackage{booktabs}

\newcommand{\dd}{\mathrm{d}}
\newcommand{\bk}{\mathbf{k}}
\newcommand{\bp}{\mathbf{p}}
\newcommand{\br}{\mathbf{r}}
\newcommand{\bR}{\mathbf{R}}
\newcommand{\rev}[1]{#1}
\newcommand{\redrev}[1]{#1}
\newcommand{\apprev}[1]{#1}
\newcommand{\epsvac}{\rev{\epsilon_0}}
\newcommand{\epsenv}{\varepsilon}
\newcommand{\epsstat}{\varepsilon_0}
\newcommand{\epsinf}{\varepsilon_\infty}
\newcommand{\sigzero}{\sigma_0}
\newcommand{\sigbulk}{\rev{\sigma_{\varepsilon}}}
\newcommand{\sigstar}{\rev{\sigma_0^{\mathrm c}}}
\newcommand{\sigt}{\sigma_{\mathrm{t}}}
\newcommand{\sigL}{\sigma_{\mathrm{L}}}
\newcommand{\alpham}{\alpha_{\mathrm{m}}}
\newcommand{\alphaeff}{\alpha_{\mathrm{eff}}}
\newcommand{\wk}{\omega_{\rm l}(k)}
\newcommand{\Etr}{\mathcal E_{\rm tr}}

\begin{document}

\title{Fr\"ohlich Bipolarons in Two-Dimensional Materials}

\author{Andrejs Kudlis}
\email{andrewkudlis@gmail.com}
\affiliation{Science Institute, University of Iceland, Dunhagi 3, IS-107 Reykjavik, Iceland}

\author{Vanik Shahnazaryan}
\email{vanikshahnazaryan@gmail.com}
\affiliation{Department of Physics and Engineering, ITMO University, St. Petersburg, 197101, Russia}
\affiliation{Department of Physics, United Arab Emirates University, P.O. Box 15551 Al-Ain, United Arab Emirates}

\author{I. Iorsh}
\affiliation{Department of Physics and Engineering, ITMO University, St. Petersburg, 197101, Russia}
\email{i.iorsh@metalab.ifmo.ru}

\author{Ivan A. Shelykh}
\affiliation{Science Institute, University of Iceland, Dunhagi 3, IS-107 Reykjavik, Iceland}

\author{Ilya V. Tokatly}
\email{ilya.tokatly@ehu.es}
\affiliation{Departamento de Pol\'imeros y Materiales Avanzados: F\'isica, Qu\'imica y Tecnolog\'ia, Universidad del Pa\'is Vasco, Avenida Tolosa 72, E-20018 San Sebasti\'an, Spain}
\affiliation{Donostia International Physics Center (DIPC), E-20018 Donostia-San Sebasti\'an, Spain}
\affiliation{IKERBASQUE, Basque Foundation for Science, Plaza Euskadi 5, 48009 Bilbao, Spain}

\date{\today}

\begin{abstract}
Motivated by the progress in the physics of two-dimensional materials and the recent two-dimensional generalization of the Fr\"ohlich model, we study the formation of bipolarons in polar monolayers. Because of very special nonlocal dielectric screening in two dimensions, this setting differs qualitatively from the conventional Fr\"ohlich model. In monolayers, (i) long wavelength LO phonons acquire a nontrivial dispersion; (ii) the Fr\"ohlich electron-phonon vertex becomes momentum-dependent and regular at small momenta; and (iii) direct repulsion between charge carriers takes the Keldysh-Rytova form. 
Using the Feynman path-integral variational approach, we show that the region in the parameter space where stable bipolarons exist is strongly modified compared to the usual quasi-two-dimensional model with dispersionless phonons. Specifically, in the most favorable limit, when the ratio $\sigzero$ of the \apprev{static polarizability to the high-frequency polarizability} tends to infinity, the lower critical coupling, sufficient for the formation of bipolarons, can be made arbitrarily small. More surprisingly, we demonstrate that no stable bipolarons can exist in the strong coupling limit: in the isolated monolayer the stability region is always confined to a finite range of coupling constants. In general, the stability region is strongly shifted toward large values of the \apprev{polarizability} ratio $\sigzero$, well beyond the parameter regimes in representative crystals, which indicates that polar monolayers do not favor bipolarons. 
\end{abstract}

\maketitle

\section{Introduction}
\label{sec:intro}

The Fr\"ohlich model provides a standard continuum setting for electron-phonon
coupling in condensed-matter physics~\cite{frohlich1950xx,devreese2009frohlich,franchini2021polarons} and a benchmark problem in mathematical
physics~\cite{lieb1958ground,donsker1983asymptotics,lieb1997exact}. The
self-localization picture of polarons was established in early
studies~\cite{landau1933electron,pekar1946local,landau1948effective}.  A single charge
carrier interacts with a dispersionless longitudinal optical (LO) phonon
field.  Once the band mass and
phonon frequency fix the energy and length scales, the problem contains
\redrev{a single dimensionless Fr\"ohlich coupling parameter} $\alpha$, which measures the
lattice-polarization interaction.  The one-polaron problem has consequently
served as a benchmark for \redrev{strong-coupling theory}~\cite{miyake1975strong},
\redrev{the all-coupling path-integral method}~\cite{feynman1955slow,Feynman-StatMech-book},
and diagrammatic Monte Carlo
methods~\cite{prokofev1998polaron}; broader developments are reviewed in
Refs.~\cite{alexandrov2010advances,devreese2016fr}.

Adding a second carrier changes the structure of the problem.  The common
polarization field produces retarded attraction, whereas the carriers repel
one another through the instantaneous Coulomb interaction.  A dispersionless
Fr\"ohlich bipolaron therefore has two independent parameters instead of one:
the coupling $\alpha$ and the relative strength of \redrev{the} direct repulsion,
$\eta$.  The latter parameter can be expressed through the ratio of
high-frequency and static bulk permittivities $\varepsilon_{\infty}$ and $\varepsilon_0$,
$\eta=\epsinf/\epsstat=\sigbulk^{-1}$.  Binding requires both a
sufficiently large $\alpha$ and a sufficiently strong lattice screening of the
repulsion.  In the most favorable limit $\eta\to0$, the Feynman variational
calculation gives critical couplings of approximately $2.90$ for the
dispersionless quasi-two-dimensional model and $6.84$ for the bulk
three-dimensional model~\cite{verbist1990stability,verbist1991large}.  At
finite $\eta$ the required coupling grows; within the harmonic strong-coupling
trial class, binding is
possible only for $\eta\lesssim0.079$~\cite{hiramoto1985interpolaron,kochetov1990multipolarons,verbist1992strong,smondyrev1995asymptotic}.

The broader interest in these bound pairs lies in their possible collective
behavior.  \redrev{If sufficiently mobile, bipolarons can behave as charged
composite bosons whose Bose--Einstein condensation can support
superconductivity}~\cite{schafroth1955superconductivity,schafroth1957quasichemical}.
\redrev{This mechanism has motivated bipolaronic scenarios for high-temperature
superconductivity}~\cite{alexandrov1981bipolaronic,emin1989large,emin1989singlet}.
Both binding and mobility are therefore essential: a pair must be stable
without becoming strongly self-trapped.

Both binding and mobility strongly depend on dimensionality, which changes the balance
between attraction, repulsion, and localization.  The Fr\"ohlich Hamiltonian, originally introduced for three-dimensional polar crystals,
\redrev{can be formally generalized to $n$ spatial
dimensions}~\cite{peeters1986ground}. Specifically, its two-dimensional version preserves the dispersionless LO phonons, while the electron-phonon coupling is related
to that of the bulk problem by scaling relations~\cite{wu1985twodpolaron,peeters1987scaling}. Physically, the Hamiltonian of this form describes a charge carrier interacting with LO phonon modes at the surface of
a three-dimensional polar crystal~\cite{Sak1972}. In the following, such a well-studied
dispersionless two-dimensional model is \redrev{referred to as}
\emph{quasi-two-dimensional}: the carriers move in a plane, while the LO field
and the dielectric response retain their bulk-like form.  Early continuum studies
formulated the two-carrier problem and its competition between lattice
polarization and direct repulsion~\cite{vinetskii1957interaction,vinetskii1961bipolar}.
The intercarrier correlations were then examined with several variational
constructions~\cite{kochetov1977twopolaron,suprun1982correlation,mukhomorov1982bipolaron,adamowski1989formation,bassani1991variational,kashirina2005correlations}.
The Feynman path-integral approach was first applied to the three-dimensional
bipolaron~\cite{hiramoto1985interpolaron} and subsequently used to map the
stability regions in two and three dimensions over broad coupling
ranges~\cite{verbist1990stability,verbist1991large}.
The \redrev{strong-coupling asymptotics were analyzed} separately~\cite{kochetov1990multipolarons,verbist1992strong,smondyrev1995asymptotic}, and
\redrev{rigorous criteria for Fr\"ohlich bipolaron and multipolaron binding were
subsequently established}~\cite{frank2010binding}.
Because the quasi-two-dimensional model is more favorable to binding than its
three-dimensional counterpart, it was applied to layered cuprates.  In
particular, characteristic lines constructed from dielectric data for
La$_2$CuO$_4$ and YBa$_2$Cu$_3$O$_7$ approach or enter the variational
stability region~\cite{verbist1990stability,devreese2009frohlich}.  \rev{The
model is appropriate for carriers confined to a plane in a layered bulk
crystal, but it does not describe the electrostatics of an isolated atomically
thin monolayer.}

Polar monolayers now allow to study the formation and stability of bipolarons in
an isolated two-dimensional dielectric environment. The key physical feature of this setting is that the \redrev{electrostatic
screening in an isolated monolayer differs qualitatively from the bulk screening.  In two
dimensions, the electronic polarizability introduces the characteristic length}
$r_\infty$.  \redrev{The corresponding screened Coulomb interaction, known as Keldysh--Rytova potential, is
logarithmic at distances shorter than} $r_\infty$ \redrev{and crosses over
to the bare $1/r$ Coulomb law at larger distances}~\cite{Rytova1967,Keldysh1979,cudazzo2011dielectric}.
\redrev{In polar crystals, the ionic polarizability introduces a second, static screening length}
$r_0$ \redrev{and makes the}
longitudinal mode dispersive: its frequency increases from the transverse value
$\omega_{\mathrm t}$ at long wavelengths to
$\omega_{\mathrm L}=\sqrt{r_0/r_\infty}\,\omega_{\mathrm t}$ at short
wavelengths.  The absence of screening at large distances strongly modifies the momentum dependence of the coupling between electrons and LO phonons in two-dimensional crystals. In particular, \redrev{the momentum-dependent Fr\"ohlich coupling parameter
vanishes linearly at small momenta making the monolayer electron--phonon vertex
regular rather than Coulomb singular at long wavelengths}~\cite{sohier2016two,sohier2017breakdown,ponce2023long,shahnazaryan2025polarons,kudlis2026allcoupling}.
First-principles calculations establish the importance of long-range
electron--phonon coupling and polaron formation in atomically thin polar
crystals~\cite{sio2022unified,sio2023polarons}.
The corresponding macroscopic single polaron theory predicts sizable couplings
for several polar monolayers, including hafnium dichalcogenides~\cite{shahnazaryan2025polarons,kudlis2026allcoupling}.
\redrev{The previous works therefore motivate a natural question: can two carriers bind in a
polar monolayer when its two-dimensional screening and dispersive phonons are
treated explicitly?}

We address this question using the Feynman path-integral variational method.
For a direct comparison, the quasi-two-dimensional reference and the monolayer
model are expressed in units defined by the large-momenta phonon
frequency $\omega_{\mathrm L}$ and the corresponding oscillator length scale $\ell_{\mathrm L}$.
We then introduce three dimensionless parameters:
$\sigzero=r_0/r_\infty$, $\sigL=\ell_{\mathrm L}/r_\infty$, and the effective
coupling $\alphaeff$.  A large $\sigzero$ means that ionic polarization greatly
increases the static screening length.  A large $\sigL$ means that the
electronic screening length is short compared to the longitudinal oscillator
length.  The momenta relevant to an extended carrier then remain in the 
long-wavelength part of the longitudinal phonon branch.
Finally, $\alphaeff$ measures the overall electron-phonon interaction in these
common longitudinal units. \redrev{Using the same units and trial action for
both models allows us to isolate how the monolayer phonon dispersion,
momentum-dependent electron--phonon coupling, and Keldysh--Rytova repulsion
affect bipolaron binding.}

With this common \redrev{parametrization}, we determine the Feynman-variational binding region in this three-parameter
space.  The result is qualitatively different from the dispersionless
reference.  \redrev{Binding is found only when the screening-length ratio
$r_0/r_\infty$ exceeds a minimum value that is far larger than the estimates
for representative polar monolayers.  Above this minimum, the variational
calculation favors a bipolaron only within a finite interval of couplings: two
separate polarons have the lower energy when the coupling is either too weak or
too strong.}  
The absence of stable bipolarons in the strong coupling limit is a special feature of the genuine two-dimensional problem, which indicates that in monolayers the formation of bipolarons is suppressed. On the other hand, we show that in the most favorable case, corresponding to the formal limit of asymptotically large $\sigma_0$ and $\sigma_L$, bipolarons remain stable at arbitrarily weak coupling. In other words, in monolayers, the lower critical coupling approaches zero, whereas in the conventional quasi-two-dimensional model it always stays finite.

The paper is organized as follows.
In Sec.~\ref{sec:model} we formulate the two-dimensional Fr\"ohlich model and its effective action.
In Sec.~\ref{sec:functional} we develop the Feynman variational functional,
introduce the longitudinal normalization, and \redrev{recover the
quasi-two-dimensional model as a limiting case}.
In Sec.~\ref{sec:results} \redrev{we present the critical-coupling curves and
relate them to material scales}. In the appendices we derive the one-polaron
reference functional and \redrev{strong-coupling asymptotics}, and
\redrev{describe the numerical procedure}.

\section{Two-dimensional Fr\"ohlich model}
\label{sec:model}

\subsection{Electrostatics and Hamiltonian}

We consider two carriers of equal mass $m$ \rev{and opposite spin} in a polar
monolayer embedded in an environment with average permittivity $\epsenv$.
\rev{We work in the spatially symmetric singlet sector.}  The  direct
interaction is the Keldysh-Rytova potential, whose two-dimensional Fourier
transform is
\begin{equation}
V_{\rm C}(k)=
\frac{e^2}{2\epsvac\epsenv\,k(1+r_\infty k)} = \frac{e^2}{2\epsvac\varepsilon_\infty(k) \,k}.
\label{eq:KR_q}
\end{equation}
Here, $e$ is the elementary charge, $\epsvac$ is the vacuum permittivity, and
$\epsenv$ is the average permittivity of the surrounding medium, and we introduce an effective high-frequency permittivity $\varepsilon_\infty(k)$.
Our Fourier convention is
$V_{\rm C}(r)=A^{-1}\sum_{\bk}V_{\rm C}(k)e^{i\bk\cdot\br}$;
here $A$ is the normalization area.
The screening length $r_\infty$ is proportional to the electronic 2D
polarizability \cite{cudazzo2011dielectric}.  Including polarizability of the lattice introduces a second length
$r_0>r_\infty$, so that the effective static and high-frequency dielectric functions, $\epsstat(k)$ and $\epsinf(k)$, become different and
take the following long-wavelength form
\begin{equation}
\epsstat(k)=\epsenv(1+r_0k),
\qquad
\epsinf(k)=\epsenv(1+r_\infty k).
\label{eq:eps_nonlocal}
\end{equation}
Thus, in 2D dielectric crystals, electrostatic screening is nonlocal and is characterized by effective momentum-dependent permittivities. The primitive electrostatic parameters of a monolayer are the two 2D
screening lengths, rather than independent bulk dielectric constants.

The frequency of the LO phonon mode and the corresponding Fr\"ohlich parameter are determined by the same
dielectric response~\cite{shahnazaryan2025polarons,kudlis2026allcoupling}.
Let $\omega_{\mathrm t}$ be the frequency of the transverse optical (TO) phonon.  In the polar monolayer the long-wavelength LO phonon becomes dispersive. At zero momentum, $k=0$, the LO and TO modes are degenerate, while at large momenta, the frequency $\wk$ of the LO phonon approaches the value of 
$\omega_{\mathrm L}=\sqrt{r_0/r_\infty}\,\omega_{\mathrm t}$. The effective Fr\"ohlich coupling also acquires a $k$-dependence. Explicitly, the LO frequency
and coupling parameter read \cite{shahnazaryan2025polarons,kudlis2026allcoupling},
\begin{align}
\wk
&=\omega_{\mathrm{t}}
\sqrt{\frac{\epsstat(k)}{\epsinf(k)}}
=\omega_{\mathrm{t}}
\sqrt{\frac{1+r_0k}{1+r_\infty k}}
=\omega_{\mathrm L}
\sqrt{\frac{r_0^{-1}+k}{r_\infty^{-1}+k}},
\label{eq:omega_l_q}
\\
\alpha(k)
&=
\frac{e^2}{4\pi\epsvac\hbar\wk}
\sqrt{\frac{m\wk}{2\hbar}}
\left[
\frac{1}{\epsinf(k)}-\frac{1}{\epsstat(k)}
\right].
\label{eq:alpha_q}
\end{align}
Quantizing the phonon field in a finite monolayer area $A$, with
$[\hat a_{\bk},\hat a_{\bk'}^\dagger]=\delta_{\bk,\bk'}$, and choosing a real
phase for each mode, the electron-phonon vertex is
\begin{equation}
V_k
=
\left[
\frac{\pi\alpha(k)\hbar}{Ak}\,
\hbar\wk
\sqrt{\frac{2\hbar\wk}{m}}
\right]^{1/2}.
\label{eq:Vq_alphaq}
\end{equation}
Remarkably, since $\alpha(k)/k$ approaches a finite value and $\wk\to\omega_{\mathrm t}$
at $k\to0$, the monolayer vertex $V_k$ remains finite in the long-wavelength limit. 

The translationally invariant two-carrier Hamiltonian is therefore
\begin{align}
\hat H_{\rm bp}
&=
\sum_{j=1}^{2}\frac{\hat{\bp}_j^2}{2m}
+V_{\rm C}(|\hat{\br}_1-\hat{\br}_2|)
+\sum_{\bk}\hbar\wk\,\hat a_{\bk}^\dagger\hat a_{\bk}
\nonumber\\
&\quad
+\sum_{\bk}
\left[
V_k\left(e^{i\bk\cdot\hat{\br}_1}+e^{i\bk\cdot\hat{\br}_2}\right)\hat a_{\bk}
+{\rm H.c.}
\right],
\label{eq:H_bp}
\end{align}
where H.c. denotes the Hermitian-conjugate term.  In what follows, we consider the 
zero-total-momentum ground state.
The conventional quasi-two-dimensional Fr\"ohlich Hamiltonian is recovered
formally when $\omega_{\rm l}(k)$ and $\alpha(k)$ are taken as momentum independent and
the direct interaction is replaced by its unscreened $1/k$ form. In the following, the latter is considered as a useful reference model, but we emphasize that it does not correspond to the electrostatic limit of an isolated polar monolayer.

\subsection{Effective action}

After the problem is reformulated in the standard thermodynamic path-integral formalism, and the phonons are integrated out, the zero-temperature Euclidean action for
carrier paths $\br_j(t)$ is obtained in the limit $\beta\to\infty$ as
\begin{equation}
S=S_{\rm kin}+S_{\rm C}+S_{\rm ph},
\label{eq:S_expansion}
\end{equation}
where $\beta=(k_{\mathrm B}T)^{-1}$.  The kinetic and instantaneous Coulomb
parts are
\begin{align}
S_{\rm kin}
&=\int_0^{\beta\hbar}\dd t\,
\sum_{j=1}^{2}\frac{m}{2}\dot{\br}_j^{\,2}(t),
\label{eq:S_kin}
\\
S_{\rm C}
&=\int_0^{\beta\hbar}\dd t\,
V_{\rm C}(|\br_1(t)-\br_2(t)|),
\label{eq:S_C}
\end{align}
and the retarded phonon contribution is
\begin{align}
S_{\rm ph}
&=-\sum_{i,j=1}^{2}\sum_{\bk}
\frac{|V_k|^2}{2\hbar}
\int_0^{\beta\hbar}\dd t
\int_0^{\beta\hbar}\dd t'\,
e^{-\wk|t-t'|}
\nonumber\\
&\quad\times
\exp\!\left\{i\bk\cdot
[\br_i(t)-\br_j(t')]\right\}.
\label{eq:S_ph}
\end{align}
The action has the same structure both for the physical monolayer and for the reference quasi-two-dimensional model. Only $\omega_{\rm l}(k)$,
$V_k$, and $V_{\rm C}(k)$ are model dependent.

\section{Feynman variational formulation}
\label{sec:functional}

In what follows we use $\omega_L$ for the units of frequency and $\ell_L = \sqrt{\hbar/(2m\omega_L)}$ for the units of length. 
Using the monolayer screening ratio $\sigzero$, for
Eq.~\eqref{eq:omega_l_q} we have
$\omega_{\mathrm L}=\sqrt{\sigzero}\,\omega_{\mathrm t}$.
For the dispersionless quasi-two-dimensional reference, $\omega_{\mathrm L}$
is identified with its constant LO frequency.

We evaluate the zero-temperature ground-state energy with the path-integral
variational method used in continuum bipolaron theory~\cite{feynman1955slow,hiramoto1985interpolaron,verbist1990stability,verbist1991large} that we adapt to our monolayer setting.
For a quadratic trial action $S_0$, the variational bound is
\begin{equation}
E_0\le E_0^{\rm trial}
+\lim_{\beta\to\infty}\frac{1}{\beta\hbar}
\langle S-S_0\rangle_0.
\label{eq:variational_bound}
\end{equation}
Here, $\langle\cdots\rangle_0$ denotes the functional averaging with the trial
action.  The quantity $E_0^{\rm trial}$ is the zero-temperature free energy of
that quadratic reference system; its explicit definition is given below after
the local representation of $S_0$ has been specified.

The local trial system contains the two physical carrier coordinates
$\br_1,\br_2$ and two auxiliary coordinates
$\boldsymbol{\xi}_1,\boldsymbol{\xi}_2$. The latter correspond to auxiliary harmonic oscillators, which mimic phonon degrees of freedom in the
trial model, and are subsequently removed by normalized
path integration.  Each auxiliary coordinate has a fictitious mass $M$.  A spring
with constant $\kappa$ connects each carrier to its own auxiliary coordinate,
whereas $\kappa'$ connects it to the auxiliary coordinate associated with the
other carrier.  The negative quadratic term with coefficient $K$ represents
the repulsive part of the trial interaction and $\bR$ is the equilibrium
carrier displacement. Thus, the five variational parameters are
$(M,\kappa,\kappa',K,R)$, where $R=|\bR|$.  With these definitions, the local
action before diagonalization is
\begin{align}
S_0
&=\int_0^{\beta\hbar}\dd t\Bigg\{
\sum_{j=1}^{2}\left[
\frac{m}{2}\dot{\br}_j^{\,2}
+\frac{M}{2}\dot{\boldsymbol{\xi}}_j^{\,2}
+\frac{\kappa}{2}|\br_j-\boldsymbol{\xi}_j|^2
\right]
\nonumber\\
&\quad+\frac{\kappa'}{2}\left(
|\br_1-\boldsymbol{\xi}_2-\bR|^2
+|\boldsymbol{\xi}_1-\br_2-\bR|^2\right)
\nonumber\\
&\quad-\frac{K}{2}|\br_1-\br_2-\bR|^2\Bigg\}.
\label{eq:S_trial_original}
\end{align}
The normalization of the auxiliary paths is included in the trial partition
function,
\begin{equation}
\begin{aligned}
Z_0&=\mathcal N_{\xi}^{-1}\int\prod_{j=1}^{2}
\mathcal D\br_j\,\mathcal D\boldsymbol{\xi}_j\,e^{-S_0/\hbar},\\
F_0&=-\frac{1}{\beta}\ln Z_0,
\qquad
E_0^{\rm trial}=\lim_{\beta\to\infty}F_0.
\end{aligned}
\label{eq:trial_free_energy}
\end{equation}
Here $\mathcal N_{\xi}$ is the product of the two \redrev{free} auxiliary path
integrals, so their artificial free energy is not added to the physical
variational bound.

Diagonalizing the quadratic form in Eq.~\eqref{eq:S_trial_original} gives four
normal coordinates in each Cartesian direction.  Translational invariance
makes one of them the free center-of-mass mode with zero frequency.  The other
three are internal modes; their positive frequencies divided by
$\omega_{\mathrm L}$ are denoted by $\Omega_1$, $\Omega_2$, and $\Omega_3$.
We follow the mode-numbering convention of Ref.~\cite{verbist1991large}.
The additional dimensionless quantity
$\nu=\sqrt{(\kappa+\kappa')/M}/\omega_{\mathrm L}$ is the frequency of the
free auxiliary oscillator entering $\mathcal N_{\xi}$, rather than a fourth
internal eigenfrequency.  Finally, $\rho=R/\ell_{\mathrm L}$ is the
dimensionless equilibrium separation.

The set $(\Omega_1,\Omega_2,\Omega_3,\nu,\rho)$ therefore contains the same
five variational degrees of freedom as $(M,\kappa,\kappa',K,R)$.  The mapping
is unique except for exchanging $\kappa$ and $\kappa'$, which only relabels two
equivalent configurations of the carrier--auxiliary springs.  A
stable quadratic form requires $\Omega_2>\Omega_3>0$,
$\Omega_2\ge\nu\ge\Omega_3$, and
$\Omega_1^2\ge\Omega_2^2+\Omega_3^2$; by definition, $\rho\ge0$.  The origin
of these conditions and the mode numbering are given in
Ref.~\cite{verbist1991large}, while the inverse map to the local parameters is
presented in Appendix~\ref{app:numerics}.

To evaluate the expectation value in the variational bound,
Eq.~\eqref{eq:variational_bound}, we need the trial
averages of the phase factors that appear explicitly in $S_{\rm ph}$ and, after
Fourier transformation, in $S_{\rm C}$.  There are two distinct cases: both
positions can belong to the same carrier or to different carriers.
Set $s=\omega_{\mathrm L}|t-t'|$.  For fixed $\bk$, functional integration over
the Gaussian trial paths gives a Gaussian in $k$.  The remaining two-dimensional
momentum integral includes an ordinary integration over the polar angle of
$\bk$.  These integrations act on different variables and are convergent for
the stable trial action, so their order may be interchanged. We first perform the
functional average and denote the subsequent angular average by an
overline.  The two factors are
\begin{align}
\overline{\left\langle e^{i\bk\cdot[\br_i(t)-\br_i(t')]}\right\rangle_0}
&=\exp\!\left[-\frac{\hbar k^2}{2m\omega_{\mathrm L}}D_{11}(s)\right],
\label{eq:D11_definition}
\\
\overline{\left\langle e^{i\bk\cdot[\br_1(t)-\br_2(t')]}\right\rangle_0}
&=J_0(kR)
\exp\!\left[-\frac{\hbar k^2}{2m\omega_{\mathrm L}}D_{12}(s)\right].
\label{eq:D12_definition}
\end{align}
The first line is the same-particle average and has zero mean displacement.
In the cross average, the two trial centers are separated by $\bR$; its angular
factor is $\overline{e^{i\bk\cdot\bR}}=J_0(kR)$, where $J_0$ is the Bessel
function of the first kind.  The remaining fluctuations define the
same-particle and cross correlators
\begin{align}
D_{11}(s)
&=\frac{\nu^2}{\Omega_1^2}\frac{s}{2}
+\frac{\Omega_1^2-\nu^2}{\Omega_1^2}
\frac{1-e^{-\Omega_1s}}{2\Omega_1}
\nonumber\\
&\quad+\frac{\Omega_2^2-\nu^2}{\Omega_2^2-\Omega_3^2}
\frac{1-e^{-\Omega_2s}}{2\Omega_2}
+\frac{\nu^2-\Omega_3^2}{\Omega_2^2-\Omega_3^2}
\frac{1-e^{-\Omega_3s}}{2\Omega_3},
\label{eq:D11}
\\
D_{12}(s)
&=\frac{\nu^2}{\Omega_1^2}\frac{s}{2}
+\frac{\Omega_1^2-\nu^2}{\Omega_1^2}
\frac{1-e^{-\Omega_1s}}{2\Omega_1}
\nonumber\\
&\quad+\frac{\Omega_2^2-\nu^2}{\Omega_2^2-\Omega_3^2}
\frac{1+e^{-\Omega_2s}}{2\Omega_2}
+\frac{\nu^2-\Omega_3^2}{\Omega_2^2-\Omega_3^2}
\frac{1+e^{-\Omega_3s}}{2\Omega_3}.
\label{eq:D12}
\end{align}
The equal-time cross width is
\begin{equation}
D_{12}(0)=
\frac{\Omega_2\Omega_3+\nu^2}
{\Omega_2\Omega_3(\Omega_2+\Omega_3)}>0.
\label{eq:D120_positive}
\end{equation}
Because $D_{12}(0)>0$, the corresponding Gaussian factor suppresses the
large-momentum contribution and keeps the direct-interaction integral finite.
After the normalized auxiliary paths have been integrated out, write the
resulting carrier-only trial action as
$S_0=S_{\rm kin}+S_{0,{\rm int}}$.  The corresponding dimensionless
contribution to the variational bound in
Eq.~\eqref{eq:variational_bound} is defined by
\begin{equation}
\hbar\omega_{\mathrm L}\Etr
=E_0^{\rm trial}
-\lim_{\beta\to\infty}\frac{1}{\beta\hbar}
\left\langle S_{0,{\rm int}}\right\rangle_0.
\label{eq:Etrial_definition}
\end{equation}
Here $E_0^{\rm trial}$ is the zero-temperature free energy defined in
Eq.~\eqref{eq:trial_free_energy}, and the second term subtracts the trial
interaction already included in $S_0$.  Evaluating this expression in terms
of the four frequency-like parameters gives
\begin{align}
\Etr
&=\Omega_1+\Omega_2+\Omega_3-2\nu
-\frac{\Omega_1^2-\nu^2}{2\Omega_1}
\nonumber\\
&\quad-\frac{1}{2}
\frac{\Omega_2(\Omega_2^2-\nu^2)
+\Omega_3(\nu^2-\Omega_3^2)}
{\Omega_2^2-\Omega_3^2}.
\label{eq:Etrial}
\end{align}

The dimensional Feynman energy separates into
\begin{equation}
E_{\rm bp}
=\hbar\omega_{\mathrm L}\Etr+E_{\rm ph}^{(11)}+E_{\rm ph}^{(12)}+E_{\rm C},
\label{eq:Ebp_expansion_dim}
\end{equation}
where the two phonon terms are the same-particle and cross contributions.
Each includes the two symmetry-equivalent index choices.  Before the momentum
sums are converted to integrals, they read
\begin{widetext}
\begin{align}
E_{\rm ph}^{(11)}
&=-2\int_0^\infty\dd u\sum_{\bk}
\frac{\pi\alpha(k)}{Ak}\,
\hbar\wk\sqrt{\frac{2\hbar\wk}{m}}\,
e^{-\wk u}
\exp\!\left[-\frac{\hbar k^2}{2m\omega_{\mathrm L}}D_{11}(u\omega_{\mathrm L})\right],
\nonumber\\
E_{\rm ph}^{(12)}
&=-2\int_0^\infty\dd u\sum_{\bk}
\frac{\pi\alpha(k)}{Ak}\,
\hbar\wk\sqrt{\frac{2\hbar\wk}{m}}\,
e^{-\wk u}J_0(kR)
\exp\!\left[-\frac{\hbar k^2}{2m\omega_{\mathrm L}}D_{12}(u\omega_{\mathrm L})\right].
\label{eq:Eph_sum}
\end{align}
\end{widetext}
The direct contribution is
\begin{equation}
E_{\rm C}=\sum_{\bk}
\frac{e^2J_0(kR)}{2A\epsvac\epsenv\,k(1+r_\infty k)}
\exp\!\left[-\frac{\hbar k^2}{2m\omega_{\mathrm L}}D_{12}(0)\right].
\label{eq:EC_sum}
\end{equation}
Equations~\eqref{eq:Ebp_expansion_dim}--\eqref{eq:EC_sum} retain the same
trial structure for a quasi-two-dimensional or monolayer Fr\"ohlich
bipolaron; the models differ only through $\omega_{\rm l}(k)$, $\alpha(k)$, and the
direct-interaction kernel.

We now express this common functional in the longitudinal units defined
above, which also makes the quasi-two-dimensional limit transparent.
Together with $\sigzero$, the other two independent dimensionless parameters are
\begin{align}
\sigL
&=\frac{\ell_{\mathrm L}}{r_\infty},
\label{eq:sigmaL}
\\
\alphaeff
&=\frac{\ell_{\mathrm L}}{a_{\mathrm B}}
\left(1-\frac{1}{\sigzero}\right)
\nonumber\\
&=\frac{e^2}{4\pi\epsvac\epsenv\hbar\omega_{\mathrm L}}
\sqrt{\frac{m\omega_{\mathrm L}}{2\hbar}}
\left(1-\frac{1}{\sigzero}\right),
\label{eq:alphaL_expanded}
\end{align}
with $a_{\mathrm B}=4\pi\epsvac\epsenv\hbar^2/(me^2)$ being the Bohr radius.  Hence
$(\sigzero,\sigL,\alphaeff)$ is a complete set of three independent parameters.
Its one-to-one relation to the transverse one-polaron parametrization used in
our previous work~\cite{kudlis2026allcoupling} is given in
Appendix~\ref{app:polaron}.

\rev{To convert Eqs.~\eqref{eq:Eph_sum} and~\eqref{eq:EC_sum} to dimensionless
momentum integrals, set}
\begin{equation}
q=k\ell_{\mathrm L},
\qquad
x_q=\frac{q}{\sigL},
\qquad
\tau=\wk u.
\label{eq:xq}
\end{equation}
\rev{The main bipolaron calculations below use $\rho=0$.  Independent checks
with $\rho$ free found no lower finite-separation state; the procedure is
summarized in Appendix~\ref{app:numerics}.}
The dimensionless phonon dispersion is
\begin{equation}
\bar\Omega_q=\frac{\wk}{\omega_{\mathrm L}}
=\sqrt{\frac{1+\sigzero x_q}{\sigzero(1+x_q)}}.
\label{eq:Omega_bar}
\end{equation}
By substituting Eqs.~\eqref{eq:eps_nonlocal} and \eqref{eq:omega_l_q} into Eq.~\eqref{eq:alpha_q}, and expressing the result in terms of the above dimensionless variables, we express the $k$-dependent Fr\"ohlich coupling in the following factorized form,
\begin{equation}
\alpha(k)=\alphaeff\frac{F(q)}{\sqrt{\bar\Omega_q}},
\qquad
F(q)=\frac{\sigzero x_q}{(1+x_q)(1+\sigzero x_q)}.
\label{eq:alpha_factorization}
\end{equation}
The corresponding dimensionless Coulomb factor is
\begin{equation}
C(q)=\frac{1}{1+x_q}.
\label{eq:Fph_FC}
\end{equation}
The function $F(q)$ is the residual momentum-dependent factor in $\alpha(k)$
after the overall coupling $\alphaeff$ and the frequency factor
$\bar\Omega_q^{-1/2}$ have been extracted.

Passing to the continuum limit according to
$\sum_{\bk}\to A(2\pi)^{-2}\int\dd^2k
=A(2\pi)^{-1}\int_0^\infty k\,\dd k$
and defining $\varepsilon_{\rm bp}=E_{\rm bp}/(\hbar\omega_{\mathrm L})$,
the final monolayer variational functional used below, takes the form
\begin{align}
\varepsilon_{\rm bp}
&=\Etr
-2\alphaeff\!\int_0^\infty\!\dd\tau\,e^{-\tau}
\int_0^\infty\!\dd q\,F(q)
e^{-q^2D_{11}(\tau/\bar\Omega_q)}
\nonumber\\
&\quad-2\alphaeff\!\int_0^\infty\!\dd\tau\,e^{-\tau}
\int_0^\infty\!\dd q\,F(q)J_0(q\rho)
e^{-q^2D_{12}(\tau/\bar\Omega_q)}
\nonumber\\
&\quad+2\alphaeff\frac{\sigzero}{\sigzero-1}
\int_0^\infty\!\dd q\,C(q)J_0(q\rho)e^{-q^2D_{12}(0)}.
\label{eq:nl2d_long_bp}
\end{align}
The binding energy is defined by comparing the bipolaron energy with the energy of two independently optimized polarons,
\begin{equation}
E_{\rm bind}=2E_{\rm pol}-E_{\rm bp}.
\label{eq:Ebind_def}
\end{equation}
Its dimensionless counterpart is
$\varepsilon_{\rm bind}=E_{\rm bind}/(\hbar\omega_{\mathrm L})$.
Positive $E_{\rm bind}$ is a binding indicator within the chosen
Feynman trial family.  Since the one- and two-polaron energies are separately
optimized upper bounds, their difference is not itself a rigorous bound on
the exact binding energy.  The one-polaron functionals are given in
Appendix~\ref{app:polaron}.
The conventional quasi-two-dimensional functional of Refs.~\cite{verbist1990stability,verbist1991large} is recovered from
Eq.~\eqref{eq:nl2d_long_bp} by setting
$\bar\Omega_q=F(q)=C(q)=1$ and identifying $\alphaeff$ with the standard Fr\"ohlich constant
$\alpha$.  In the quasi-two-dimensional reference, both the electron-phonon coupling and the direct Coulomb repulsion are determined by the bulk permittivities. In particular, the coefficient in the direct repulsion term is expressed in terms of the ratio $\sigbulk=\epsstat/\epsinf$, as follows,
$2\alpha\sigbulk/(\sigbulk-1)$. In other words, the direct repulsion term for the quasi-two-dimensional model is obtained from the last term in Eq.~\eqref{eq:nl2d_long_bp} by replacing the ratio $\sigzero=r_0/r_\infty$ of the 2D screening lengths by the ratio $\sigbulk=\epsstat/\epsinf$ of the bulk permittivities. 

The three replacements in the integral kernels in Eq.~\eqref{eq:nl2d_long_bp}  have distinct significance. Setting
$\bar\Omega_q=1$ removes phonon dispersion from the retardation arguments,
while $C(q)=1$ removes the high-frequency Keldysh--Rytova denominator.  The
replacement $F(q)=1$ is more consequential. For the monolayer, at finite $\sigzero$,
\begin{equation}
F(q)\simeq\frac{\sigzero}{\sigL}q,
\qquad
C(q)\simeq1
\qquad(q\to0)\rev{.}
\label{eq:small_q}
\end{equation}
Thus, the phonon kernel vanishes at small momenta in the monolayer, whereas
it remains finite in the quasi-two-dimensional model.  Setting $F(q)=1$
removes this physical difference; the quasi-two-dimensional model is not
obtained simply by taking $q\to0$ in the monolayer equations.
\section{Results}
\label{sec:results}

\subsection{Dispersionless quasi-two-dimensional dielectric line}
\label{subsec:vpd_results}

The dispersionless quasi-two-dimensional model provides a natural reference for the
monolayer calculation. Let us briefly describe its behavior. The most favorable for the formation of bipolarons is the limit $\sigbulk\to\infty$ in which the direct repulsion $2\alpha\sigbulk/(\sigbulk-1)$ is minimal. In this limit, stable \apprev{bipolarons} are formed if the coupling exceeds the critical value of
\begin{equation}
\alpha_{\mathrm c}(\infty)\simeq2.90.
\label{eq:vpd_alpha_inf}
\end{equation}
This minimal critical coupling was reported in the original numerical Feynman phase
diagram~\cite{verbist1990stability} and discussed in detail in~\cite{verbist1991large}.

\begin{figure}[t]
\centering
\includegraphics[width=\columnwidth]{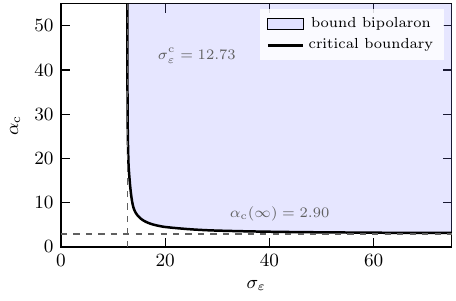}
\caption{Critical coupling of the dispersionless quasi-two-dimensional model
\redrev{under the dielectric constraint specified in the text}.  The blue area
\redrev{shows where the variational calculation favors a bound bipolaron}
within the displayed
parameter range.  The solid curve shows the calculated
critical boundary.  The horizontal dashed line is
$\alpha_{\mathrm c}(\infty)\simeq2.90$; the vertical dashed line marks the
strong-coupling end point \rev{$\sigbulk^{\mathrm c}\simeq12.73$}.}
\label{fig:q2d-critical}
\end{figure}

Reducing \rev{$\sigbulk$} strengthens the repulsion relative to the phonon-mediated
attraction, so that critical coupling rises and finally diverges at
\begin{equation}
\rev{\sigbulk^{\mathrm c}\simeq12.73.}
\label{eq:vpd_sigma_star}
\end{equation}
The same critical ratio of permittivities is also obtained from the strong-coupling
analysis for a bulk system ~\cite{hiramoto1985interpolaron,kochetov1990multipolarons,smondyrev1995asymptotic}.
Within the harmonic Feynman trial class, the coincidence of $\sigbulk^{\mathrm c}$ for three-dimensional and quasi-two-dimensional Fr\"ohlich  models is a manifestation of the standard scaling relations~\cite{peeters1987scaling,verbist1992strong}.  
Figure~\ref{fig:q2d-critical} displays the numerical critical line and both
asymptotes. We emphasize that for the dielectric ratio \rev{$\sigbulk<\sigbulk^{\mathrm c}$}, no finite coupling binds the
quasi-two-dimensional bipolaron within this variational model.

\begin{figure}[!t]
\centering
\includegraphics[width=\columnwidth]{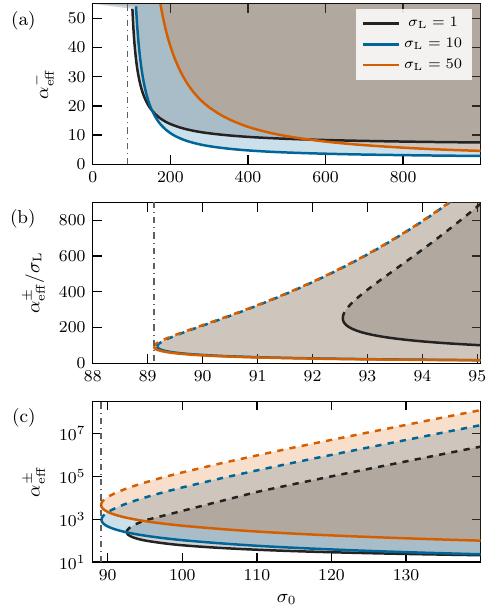}
\caption{Binding boundaries of the dispersive monolayer model for three
representative longitudinal length ratios.  (a) Lower critical coupling over
the displayed $\sigzero$ range.  (b) Both boundaries near their meeting
point, shown as $\alpha_{\mathrm{eff}}^\pm/\sigL$ so their smooth meeting can be
resolved on a linear scale.  (c) The same binding windows over a broader range;
the logarithmic vertical scale displays the rapidly growing upper boundary.
Solid and dashed lines denote the lower and upper boundaries, respectively.
The colored shading marks the displayed part of each binding region: above
the lower boundary in panel (a), and between the matching lower and upper
boundaries in panels (b) and (c).  Binding is identified by
$\varepsilon_{\rm bind}>0$ within the searched Feynman trial class.  The
dash-dotted vertical line marks the large-$\sigL$ limit
\rev{$\sigstar=89.11$}.  \rev{In panel (a), the shading indicates only the side
on which binding first appears; it should not be read as extending beyond the
upper boundary shown in panels (b) and (c).}}
\label{fig:monolayer-phase}
\end{figure}

\subsection{Dispersive monolayer phase diagram}
\label{subsec:nl2d_results}

As it has been indicated, the monolayer is specified by the three independent parameters
$\sigzero$, $\sigL$, and $\alphaeff$.  For each fixed pair $(\sigzero,\sigL)$ we
reoptimize the one-polaron energy and the bipolaron energy at
every coupling strength $\alphaeff$.  A point is classified as bound when \rev{the variational indicator in
Eq.~\eqref{eq:Ebind_def} is positive}. By performing calculations according to the numerical procedure described in Appendix~\ref{app:numerics}, we find that  whenever binding occurs, it corresponds to a finite
interval of coupling constants
\begin{equation}
\alpha_{\mathrm{eff}}^-<\alphaeff<\alpha_{\mathrm{eff}}^+,
\label{eq:binding_interval}
\end{equation}
where $\alpha_{\mathrm{eff}}^-$ is the lower critical coupling and
$\alpha_{\mathrm{eff}}^+$ is the upper one. Therefore, in contrast to the quasi-two-dimensional case in the isolated monolayer, no stable bipolaron state exists in the strong coupling limit $\alphaeff\to\infty$ independently of the other two parameters.

Figure~\ref{fig:monolayer-phase}(a) shows that the lower critical coupling decreases as
$\sigzero$ is increased, and it depends strongly on $\sigL$.  At any finite
$\sigL$, its large-$\sigzero$
limit remains nonzero.  The curves need not be ordered by $\sigL$ at finite
$\sigzero$: increasing $\sigL$ lowers the eventual large-$\sigzero$ limit, but
also delays the approach of $\alphaeff^-$ to its limiting value, so the curves cross within the displayed
range.  However, asymptotically, the lower critical coupling approaches zero in
the ordered fashion if the limit $\sigzero\to\infty$ is taken first and then followed by $\sigL\to\infty$. The asymptotic analysis in
Appendix~\ref{app:zero_corner} shows that at $\sigzero\to\infty$ and large $\sigL$, the critical coupling $\alphaeff^-$ takes the form,
\begin{equation}
\alpha_{\mathrm{eff}}^-
=a_{\rm c}\sigL^{-1/3}\rev{+o(\sigL^{-1/3})},
\qquad a_{\rm c}\simeq3.4567.
\label{eq:zero_corner_result}
\end{equation}
\rev{Direct finite-$\sigL$ minimization confirms this scaling, as shown in
Table~\ref{tab:zero-corner-convergence}.} 

Therefore, in the monolayer, the minimal critical coupling goes to zero, which is qualitatively different from the quasi-two-dimensional model, where $\alpha_c$ is finite and given by Eq.~\eqref{eq:vpd_alpha_inf}. In  the asymptotic regime described by Eq.~\eqref{eq:zero_corner_result} the bipolaron physics is dominated by the phonon dispersion, and this determines the main difference from dispersionless Fr\"ohlich models. It is worth noting that this is a singular soft-phonon limit. The binding scale also tends to zero, and this regime does not correspond to ordinary weak coupling at fixed material parameters.

\begin{table}[!b]
\caption{Numerical check of the asymptotic law in
Eq.~\eqref{eq:zero_corner_result} after taking $\sigzero\to\infty$.  The lower
critical coupling $\alpha_{\mathrm{eff}}^-$ decreases with $\sigL$, while the
scaled product $\alpha_{\mathrm{eff}}^-\sigL^{1/3}$ approaches
$a_{\rm c}\simeq3.4567$.}
\label{tab:zero-corner-convergence}
\centering
\footnotesize
\setlength{\tabcolsep}{0.8pt}
\begin{ruledtabular}
\begin{tabular}{c|rrrrrrr}
$\sigL$ & $1$ & $2$ & $5$ & $10$ & $50$ & $200$ & $1000$\\
\hline
$\alpha_{\mathrm{eff}}^-$
& $6.1158$ & $3.8534$ & $2.3360$ & $1.7161$ & $0.9467$ & $0.5919$ & $0.3457$\\
$\alpha_{\mathrm{eff}}^-\sigL^{1/3}$
& $6.1158$ & $4.8550$ & $3.9946$ & $3.6971$ & $3.4878$ & $3.4617$ & $3.4573$
\end{tabular}
\end{ruledtabular}
\end{table}

\begin{table*}[!t]
\caption{Screening and coupling parameters for representative polar monolayers. Parameters $\sigL$ and $\alphaeff$ depend on the effective mass of charge carriers and take values $\sigL^{\rm e}$, $\alphaeff^{\rm e}$ and $\sigL^{\rm h}$, $\alphaeff^{\rm h}$  for electrons and holes, respectively. For completeness, in the last columns, we show alternative \apprev{dimensionless} parameters,
$\sigt=r_{\mathrm t}/r_\infty$ and
$\alpham=(r_{\mathrm t}/a_{\mathrm B})(\sqrt{\sigzero}-1)/(\sqrt{\sigzero}+1)$, used to characterize polar monolayers in Ref.~\cite{kudlis2026allcoupling}. The relation between the two parameterizations is discussed in Appendix~A1.}
\label{tab:materials_sigma0}
\centering
\small
\begin{ruledtabular}
\begin{tabular}{l|c|c|c|cc|cc|cc|cc}
Material
& $r_\infty$ (nm)
& $r_0$ (nm)
& $\sigzero$
& $\sigL^{\mathrm e}$
& $\sigL^{\mathrm h}$
& $\alphaeff^{\mathrm e}$
& $\alphaeff^{\mathrm h}$
& $\sigt^{\mathrm e}$
& $\sigt^{\mathrm h}$
& $\alpham^{\mathrm e}$
& $\alpham^{\mathrm h}$\\
\hline
h-BN  & 0.78 & 1.08 & 1.38 & 0.61 & 0.69 & 2.06 & 1.82 & 0.66 & 0.75 & 0.65 & 0.58\\
GaN   & 0.76 & 1.11 & 1.47 & 1.77 & 0.75 & 1.93 & 4.58 & 1.95 & 0.82 & 0.64 & 1.51\\
AlN   & 0.47 & 0.76 & 1.64 & 1.90 & 1.11 & 3.34 & 5.70 & 2.16 & 1.26 & 1.19 & 2.03\\
HfSe$_2$ & 4.25 & 21.8 & 5.12 & 0.68 & 0.61 & 7.95 & 8.99 & 1.03 & 0.91 & 5.75 & 6.50\\
HfS$_2$ & 2.97 & 14.0 & 4.70 & 0.68 & 0.50 & 7.25 & 9.81 & 1.00 & 0.74 & 5.00 & 6.77\\
ZrS$_2$ & 2.83 & 10.6 & 3.76 & 0.52 & 0.56 & 6.27 & 5.75 & 0.72 & 0.78 & 3.80 & 3.48
\end{tabular}
\end{ruledtabular}
\end{table*}

By contrast, the upper boundary has a different origin and can be understood
analytically from the large-$q$ behavior of the phonon and Coulomb terms. The phonon form factor $F(q)$ defined in Eq.~\eqref{eq:alpha_factorization} can be represented as follows,
\begin{equation}
F(q)
=\frac{\sigzero}{\sigzero-1}
\left(\frac{1}{1+x_q}-\frac{1}{1+\sigzero x_q}\right).
\label{eq:Fph_partial_main}
\end{equation}
Thus, both $F$ and $C$ behave as $\sigL/q$ at large momentum.
The quasi-two-dimensional model instead has constant radial form factors, which
makes its strong-coupling energy quadratic in the coupling. For monolayer energy functional Eq.~\eqref{eq:nl2d_long_bp}, the $1/q$ decay of the Coulomb and the phonon form factors gives momentum integrals of the form $\int dq/q$ with the logarithmic divergences cut off by the Gaussian factors. Upon the minimization, the logarithmic behavior of the integrals translates to the logarithmic dependence of
the interaction energy on the coupling. This eventually generates the upper binding boundary at finite $\sigzero$.

By separately optimizing the one-polaron and \apprev{localized bipolaron} energies, and considering the leading terms in the strong-coupling limit of $\alphaeff\to\infty$ at fixed
$\sigL$ and $\sigzero$, we find,
\begin{equation}
\frac{\varepsilon_{\rm bind}}{\alphaeff\sigL}
=-\frac{1}{\sigzero-1}\ln\!\left(\frac{\alphaeff}{\sigL}\right)
+\mathcal B(\sigzero)+o(1).
\label{eq:Ebind_compact_main}
\end{equation}
where the explicit form of \apprev{$\mathcal B(\sigzero)$} is presented in
Appendix~\ref{app:nl2d_asymptotic} together with details of the derivation. Obviously, at finite $\sigzero$, the
logarithmic term in Eq.~\eqref{eq:Ebind_compact_main} becomes increasingly
negative with the increase of $\alphaeff$, and at sufficiently large coupling, this term always wins, signifying instability of bipolaron state. For $\sigzero\gg 1$ the function $\mathcal{B}(\sigzero)$ approaches a finite value $\mathcal B(\sigzero)\to\ln(32/27)>0$, see Appendix~\ref{app:nl2d_asymptotic}~3. Therefore, with the increase of $\sigzero$ the upper critical coupling grows exponentially, $\alphaeff^+\approx\sigL e^{\sigzero\mathcal{B}}$, as can be seen in Fig.~2~(b-c). 

By comparing the phase diagrams in Figs.~1 and 2, we also observe an overall shift of the bipolaron stability region toward larger \apprev{polarizability} ratios in the monolayer case. 
\rev{The shift of the lower boundary reflects the small-momentum mismatch in different interaction channels, see
Eq.~\eqref{eq:small_q}. Because of the absence of screening at large distances the retarded phonon-mediated attraction is
suppressed while the direct repulsion remains finite. The logarithmic
large-momentum behavior discussed above produces the upper boundary.  Together,
these mechanisms shift the stability region and confine binding to a finite coupling interval, making the monolayer less
favorable for bipolarons than in the dispersionless quasi-two-dimensional model.}

When $\sigzero$ is reduced, the lower and upper critical curves meet smoothly, as shown in
Fig.~\ref{fig:monolayer-phase}(b)-(c). The horizontal position $\sigzero^{\mathrm c}(\sigL)$ of the meeting point moves
from about $92.55$ at $\sigL=1$ toward a common limit at large $\sigL$. \apprev{The endpoint intervals obtained at large finite $\sigL$ overlap near $89.11$, consistently with this limiting value.} A careful asymptotic analysis of the large $\sigL$ regime
performed in Appendix~\ref{app:nl2d_boundary_meeting} identifies the meeting point of the upper and lower boundaries as
\begin{equation}
\sigstar\simeq89.11,
\qquad
\alphaeff^{\mathrm c}
\simeq 91.76\,\sigL.
\label{eq:nl2d_boundary_meeting_result}
\end{equation}
Thus, at large $\sigL$, the critical coupling at the meeting point scales linearly with $\sigL$. As shown in Fig.~\ref{fig:monolayer-phase}(b) and confirmed analytically in  Appendix~\ref{app:nl2d_boundary_meeting}, this scaling holds true well beyond the meeting point. The critical value $\sigstar\simeq89.11$ defines the minimal \apprev{polarizability} ratio $\sigzero$ at which stable bipolarons can exist. It is worth noting that this value is about 7 times larger than the critical ratio $\sigbulk^{\mathrm c}\simeq12.73$ for the quasi-two-dimensional model. 

After identifying the set of control parameters and the stability region of bipolaron states in the continuum model of polar monolayers, we briefly discuss the relation of these results to real 2D materials. Table~\ref{tab:materials_sigma0} presents values of material parameters for a set of polar monolayers, ranging from weakly and moderately coupled systems, such as h-BN or GaN, to Hf-based materials with relatively strong coupling. As we can see, even in the most favorable cases of holes in HfSe$_2$ and HfS$_2$, the ratio of screening lengths $\sigzero$ is more than an order of magnitude of a critical value required for the formation of bipolarons. Therefore, even in materials where the binding energy of a singe polaron is quite sizable, about 150~meV for Hf-compounds \cite{kudlis2026allcoupling}, the polarons do not form stable two-particle bound states.

\FloatBarrier
\section{Conclusions}
\label{sec:conclusions}

We have studied the formation of Fr\"ohlich bipolarons in an isolated polar monolayer
using the Feynman path-integral variational approach. The results are contrasted to those for the quasi-two-dimensional Fr\"ohlich model that describes coupling to LO phonons at the surface of 3D insulators \cite{Sak1972} and is commonly used to address polaron physics in reduced dimensions \cite{peeters1986ground,wu1985twodpolaron,verbist1990stability,verbist1991large}. Compared to the quasi-two-dimensional reference, the polaron effects in isolated monolayers are determined by a combination of dispersive LO phonons with momentum-dependent electron-phonon
coupling and Keldysh-Rytova repulsion between electrons. One of the most critical features in 2D insulators is the absence of screening at large distances. This removes the Coulomb singularity in the phonon-mediated attraction and suppresses the binding of polarons. In this work, we study and characterize these effects quantitatively.
We show that there is a qualitative modification of the bipolaron stability region in the parameter space. Firstly, in the regime corresponding to an ordered limit $\sigzero\to\infty$ followed by $\sigL\to\infty$, the lower critical coupling can be made arbitrarily weak, whereas in the standard Fr\"ohlich model it is always finite. The existence of this regime is a direct manifestation of the phonon dispersion in genuine 2D crystals. Another qualitative feature of the bipolaron problem in the monolayer is the existence of the upper critical coupling, which implies the absence of stable bipolarons in the strong coupling limit. This seemingly counterintuitive result can be traced back to the logarithmic behavior of the Keldysh-Rytova potential at short distances that become relevant at strong coupling. Finally, the suppression of the electron-phonon coupling at small momenta leads to an overall shift of the stability region toward large values of the \apprev{polarizability} ratio $\sigzero$, well outside the parameter range realized in currently available 2D materials.   

All quantitative results of this work are obtained using the Feynman path-integral variational principle within the harmonic variational class. From the previous experience with similar problems, one usually expects this method to produce sufficiently accurate results. Moreover, our analytic and physical arguments regarding the qualitative shape of the stability region should be valid generically. Nonetheless, it would be interesting to rigorously confirm our conclusions in a way similar to the rigorous treatment of the standard Fr\"ohlich bipolaron problem \cite{frank2010binding}.

\begin{acknowledgments}
The work of A.K. is supported by the Icelandic Research Fund (Ranns\'oknasj\'o{\dh}ur, Grant No.~2410550).
V.S. acknowledges the support of ``Basis'' Foundation (Project No.~25-1-3-11-1).
IVT acknowledges support from the Spanish MCIN/AEI/10.13039/501100011033 through the project PID2023-148225NB-C32, and the Basque Government (Grant No.~IT1453-22).
\apprev{We acknowledge access to the Elja computing cluster through the
Icelandic Research e-Infrastructure (IREI). These resources are provided by
the University of Iceland's Division of Information Technology and funded by
the Icelandic Infrastructure Fund, administered by Rann\'is (the Icelandic
Centre for Research).}
\end{acknowledgments}

\appendix
\raggedbottom

\section{Single-polaron counterparts}
\label{app:polaron}

\subsection{Connection between different \apprev{parametrizations}}

In general the physics of the two-dimensional Fr\"ohlich model is controlled by three independent \apprev{dimensionless} parameters. The original papers \cite{shahnazaryan2025polarons,kudlis2026allcoupling}, where this model is introduced, adopt \apprev{a} parametrization which uses the frequency \apprev{$\omega_{\mathrm t}$} of TO phonons and the corresponding oscillator length scale,
\begin{equation}
r_{\mathrm t}=\sqrt{\frac{\hbar}{2m\omega_{\mathrm t}}}\rev{,}
\label{eq:length_scales_native}
\end{equation}
as the basic frequency and the length units. The corresponding three dimensionless parameters are
\begin{equation}
\sigzero=\frac{r_0}{r_\infty},
\qquad
\sigt=\frac{r_{\mathrm t}}{r_\infty},
\qquad
\alpham=\frac{r_{\mathrm t}}{a_{\mathrm B}}
\frac{\sqrt{\sigzero}-1}{\sqrt{\sigzero}+1}.
\label{eq:sigmas_native}
\end{equation}
In this set, $\alpham$ provides a strict upper bound \apprev{on} the momentum-dependent coupling $\alpha(k)$. \apprev{The parameters $\sigzero$, $\sigL$, and $\alphaeff$ used here, defined in Eqs.~\eqref{eq:sigmaL}--\eqref{eq:alphaL_expanded}, are related to this parametrization by}
\begin{equation}
\sigL=\frac{\sigt}{\sigzero^{1/4}},
\qquad
\alphaeff=\alpham
\frac{(\sqrt{\sigzero}+1)^2}{\sigzero^{5/4}}.
\label{eq:alphaL_map}
\end{equation}
The inverse transformation reads
\begin{equation}
\sigt=\sigL\sigzero^{1/4},
\qquad
\alpham=\alphaeff
\frac{\sigzero^{5/4}}{(\sqrt{\sigzero}+1)^2}.
\label{eq:inverse_map}
\end{equation}

\subsection{One-polaron energy}

The one-polaron Feynman trial action is parametrized by two frequencies $v>w>0$.
\rev{The dimensionless displacement function entering its Gaussian phase
factor, analogous to $D_{11}$, is}
\begin{equation}
D_{\rm pol}(s)
=
\frac{w^2}{v^2}s
+\frac{v^2-w^2}{v^3}\left(1-e^{-vs}\right),
\label{eq:Dpol}
\end{equation}
\rev{and its trial-energy contribution is}
\begin{equation}
\mathcal E_{\rm tr}^{\rm pol}=\frac{(v-w)^2}{2v}.
\label{eq:Etr_pol}
\end{equation}
For the dispersionless quasi-two-dimensional reference problem, the corresponding
one-polaron functional is
\begin{align}
\varepsilon_{\rm pol}
&=
\mathcal E_{\rm tr}^{\rm pol}
-\alpha
\int_0^\infty\dd\tau\,e^{-\tau}
\int_0^\infty\dd q\,
e^{-q^2D_{\rm pol}(\tau)}.
\label{eq:Epol_VPD}
\end{align}
The \apprev{monolayer one-polaron functional} used in the binding-energy
comparison is
\begin{align}
\varepsilon_{\rm pol}
&=
\mathcal E_{\rm tr}^{\rm pol}
-\alphaeff
\int_0^\infty\dd\tau\,e^{-\tau}
\int_0^\infty\dd q\,
F(q)
\nonumber\\
&\quad\times
e^{-q^2D_{\rm pol}(\tau/\bar\Omega_q)}.
\label{eq:Epol_2D}
\end{align}
Equation~\eqref{eq:Epol_VPD} is recovered from Eq.~\eqref{eq:Epol_2D} by
setting $F=1$, $\bar\Omega_q=1$, and replacing $\alphaeff$ by the
quasi-two-dimensional coupling $\alpha$.
Both the one-polaron and the bipolaron energies are minimized independently
within the same model before forming Eq.~\eqref{eq:Ebind_def}.

\section{Vanishing lower threshold at large screening lengths}
\label{app:zero_corner}

This appendix derives Eq.~\eqref{eq:zero_corner_result}.  \apprev{We first take}
$\sigzero\to\infty$ in the full monolayer
functional and then take $\sigL\to\infty$.  At the first step,
\begin{equation}
 \bar\Omega_q^2=\frac{q/\sigL}{1+q/\sigL},
 \qquad
 F(q)=C(q)=\frac{1}{1+q/\sigL}.
\label{eq:zero_ordered_kernels}
\end{equation}
\rev{These are the kernels used below; both the phonon dispersion and the
momentum denominator are retained in the subsequent large-$\sigL$ scaling.}

\apprev{The characteristic momentum, trial frequencies, and critical coupling
selected by minimization vary with $\sigL$.  The algebraic dependence of the
kernels on $q/\sigL$ suggests a power-law scaling of these quantities.
Its exponents follow by retaining a finite time dependence and Gaussian
exponent $q^2D$, while keeping the trial and interaction energies comparable.}

Accordingly, write the leading behavior as
\begin{equation}
 q=\sigL^{-\gamma}p,
 \qquad
 \alphaeff=a\sigL^{-\chi},
 \qquad
 \mathcal V=\sigL^{-\zeta}\widehat{\mathcal V},
\label{eq:zero_corner_general_powers}
\end{equation}
where $\mathcal V=(v,w,\Omega_1,\Omega_2,\Omega_3,\nu)$ is shorthand for the
two independently optimized sets of trial frequencies.  The hatted
frequencies and $a$ remain finite, while $p=O(1)$ labels the rescaled momentum
region.  Equation~\eqref{eq:zero_ordered_kernels}
gives
\begin{equation}
 \bar\Omega_q^2\sim\frac q{\sigL}=\sigL^{-(1+\gamma)}p.
\label{eq:zero_corner_dispersion_balance}
\end{equation}
The correlators contain the trial frequencies multiplied by
$\tau/\bar\Omega_q$.  A nontrivial time dependence therefore requires the
trial and phonon frequencies to have the same scale,
\begin{equation}
 \zeta=\frac{1+\gamma}{2}.
\label{eq:zero_corner_time_balance}
\end{equation}
For $\tau=O(1)$ and $p=O(1)$, Eqs.~\eqref{eq:Dpol}, \eqref{eq:D11}, and
\eqref{eq:D12} imply $D_j=O(\sigL^\zeta)$, where $D_j$ denotes any of these
three correlators.  Hence
$q^2D_j=O(\sigL^{\zeta-2\gamma})$, and a finite Gaussian exponent requires
$\zeta=2\gamma$.  Together with Eq.~\eqref{eq:zero_corner_time_balance}, this gives
\begin{equation}
 \gamma=\frac13,
 \qquad
 \zeta=\frac23.
\label{eq:zero_corner_q_frequency_powers}
\end{equation}
Finally, the radial measure contributes $\dd q=O(\sigL^{-\gamma})$.  The
remaining scaled kernels and correlators are of order unity, so the
interaction energy is $O(\sigL^{-\chi-\gamma})$, whereas the trial energy is
$O(\sigL^{-\zeta})$.  Their balance fixes
\begin{equation}
 \chi+\gamma=\zeta,
 \qquad
 \chi=\frac13.
\label{eq:zero_corner_coupling_power}
\end{equation}
These three balances therefore give
\begin{equation}
 q=\sigL^{-1/3}p,\qquad
 \alphaeff=a\sigL^{-1/3},\qquad
\mathcal V=\sigL^{-2/3}\widehat{\mathcal V}.
\label{eq:zero_corner_scaling}
\end{equation}
\rev{As in the main analysis, we set $\rho=0$, so the universal functional
below contains no separation variable.}

Under the scaling in Eq.~\eqref{eq:zero_corner_scaling}, the kernels in
Eq.~\eqref{eq:zero_ordered_kernels} become
\begin{align}
 \sigL^{4/3}\bar\Omega_{\sigL^{-1/3}p}^2
 &=\frac{p}{1+\sigL^{-4/3}p}\longrightarrow p,
 \nonumber\\
 F(\sigL^{-1/3}p)&=C(\sigL^{-1/3}p)
 =\frac{1}{1+\sigL^{-4/3}p}\longrightarrow1.
\label{eq:zero_finite_kernels}
\end{align}
Direct substitution of the scaled frequencies in the correlators yields
\begin{equation}
 D_j\!\left(\frac{\tau}{\bar\Omega_{\sigL^{-1/3}p}}\right)
 =\sigL^{2/3}\widehat D_j\!\left(
 \frac{\tau}{\sqrt p}\right)+o(\sigL^{2/3}).
\label{eq:zero_scaled_D}
\end{equation}
\rev{Here $\widehat D_j$ is obtained from the corresponding correlator by
replacing every trial frequency by its hatted value.}  Thus
$q^2D_j=p^2\widehat D_j+o(1)$ and every term \rev{in the two variational
energies} has the common factor $\sigL^{-2/3}$.

Taking $\sigL\to\infty$ and dropping hats from the finite scaled frequencies gives
the universal one-polaron and bipolaron functionals.  Thus, in the remainder of
this appendix, $v,w,\Omega_i$, and $\nu$ denote the finite scaled frequencies in
Eq.~\eqref{eq:zero_corner_scaling}, rather than the original frequencies.
\begin{align}
 \apprev{\mathcal E_{\rm pol}(a)}
 &\apprev{{}=\mathcal E_{\rm tr}^{\rm pol}}
 \nonumber\\
 &\apprev{\quad-a\int_0^\infty\!\dd\tau\,e^{-\tau}
 \int_0^\infty\!\dd p\,
 e^{-p^2D_{\rm pol}(\tau/\sqrt{p})},}
\label{eq:zero_corner_pol}\displaybreak[0]\\
 \mathcal E_{\rm bp}(a)
 &=\Etr
 -2a\int_0^\infty\!\dd\tau\,e^{-\tau}
 \int_0^\infty\!\dd p\,
 e^{-p^2D_{11}(\tau/\sqrt{p})}
 \nonumber\\
 &\quad
 -2a\int_0^\infty\!\dd\tau\,e^{-\tau}
 \int_0^\infty\!\dd p\,
 e^{-p^2D_{12}(\tau/\sqrt{p})}
 \nonumber\\
 &\quad+2a\int_0^\infty\!\dd p\,e^{-p^2D_{12}(0)}.
\label{eq:zero_corner_bp}
\end{align}
The original energies obey
\begin{align}
 \varepsilon_{\rm pol}
 &=\sigL^{-2/3}\mathcal E_{\rm pol}+o(\sigL^{-2/3}),
 \nonumber\\
 \varepsilon_{\rm bp}
 &=\sigL^{-2/3}\mathcal E_{\rm bp}+o(\sigL^{-2/3}).
\label{eq:zero_corner_energy}
\end{align}

Let $\mathcal E_{\rm pol}^{\min}(a)$ and $\mathcal E_{\rm bp}^{\min}(a)$ denote
the independent minima of Eqs.~\eqref{eq:zero_corner_pol}
and~\eqref{eq:zero_corner_bp}, respectively.  The first minimum is taken over
$v>w>0$; the second is taken over $\Omega_i$ and $\nu$
subject to the stability conditions stated below
Eq.~\eqref{eq:S_trial_original}, with $\rho=0$.
\apprev{For the threshold search the localized solution is followed through
the crossing with two polarons; the treatment of dissociation is described
in Appendix~\ref{app:numerics}.}  Equations~\eqref{eq:zero_corner_energy}
then give
\begin{equation}
 \varepsilon_{\rm bind}
 =\sigL^{-2/3}\!\left[
 2\mathcal E_{\rm pol}^{\min}(a)-\mathcal E_{\rm bp}^{\min}(a)
 \right]+o(\sigL^{-2/3}).
\label{eq:zero_corner_binding}
\end{equation}
\apprev{At leading order, the positive prefactor leaves the crossing condition
unchanged:}
$2\mathcal E_{\rm pol}^{\min}(a_{\rm c})=
\mathcal E_{\rm bp}^{\min}(a_{\rm c})$.
\apprev{The numerical procedure and quadrature sizes are specified in
Appendix~\ref{app:numerics}.  Three successive refinements give
$3.45671914$, $3.45669678$, and $3.45669488$; the last change is
$1.9\times10^{-6}$.  An uncertainty of $10^{-5}$ gives}
\begin{equation}
 \rev{a_{\rm c}=3.45670(1).}
\label{eq:zero_corner_ac}
\end{equation}
At this threshold, the optimized scaled frequencies are
$v=9.116$, $w=2.418$, $\Omega_1=11.238$, $\Omega_2=6.991$,
$\Omega_3=0.814$, and $\nu=2.083$.  Their finite positive values \apprev{are
consistent with the scaling in} Eq.~\eqref{eq:zero_corner_scaling}.
\rev{Combining Eqs.~\eqref{eq:zero_corner_scaling} and
\eqref{eq:zero_corner_ac} yields Eq.~\eqref{eq:zero_corner_result}, including
its exponent and coefficient.}  The finite-$\sigL$ calculations summarized in
Table~\ref{tab:zero-corner-convergence} confirm convergence to
Eq.~\eqref{eq:zero_corner_ac}.  This coefficient applies to the ordered limit
stated at the beginning of the appendix.  If $\sigzero$ and $\sigL$ instead
increase simultaneously along a fixed relation, momentum-dependent factors
removed by the first limit can remain after rescaling and change the limiting
coefficient.

\section{High-coupling boundary of the monolayer model}
\label{app:nl2d_asymptotic}

This appendix derives the high-coupling expansion
\eqref{eq:Ebind_compact_main} and explains why a finite upper binding boundary
appears in the monolayer model.  We take $\alphaeff\to\infty$ at fixed
$\sigL>0$ and finite $\sigzero>1$, and set $\rho=0$.
The coupling-dependent strong-coupling scale is
\begin{equation}
g=\alphaeff\sigL.
\label{eq:g_nl2d}
\end{equation}

\subsection{Strong-coupling scale}

\rev{The standard strong-coupling hierarchy of the harmonic
Feynman trial action has two large frequencies, $\Omega_1$ and $\Omega_2$,
while $\Omega_3$ and $\nu$ remain finite~\cite{smondyrev1995asymptotic}.
\apprev{Their dependence on $g$ follows from the balance between localization
and interaction energies.}  We use the dimensionless
imaginary-time separation $s=\omega_{\mathrm L}|t-t'|$, defined before
Eqs.~\eqref{eq:D11_definition}--\eqref{eq:D12_definition}; in the phonon
functional, $s=\tau/\bar\Omega_q$.  When the two large frequencies have the
same asymptotic order,
direct substitution in Eqs.~\eqref{eq:D11}--\eqref{eq:D12} gives
that both $D_{11}$ and $D_{12}$ are inversely proportional to this common
frequency scale at every fixed $s>0$.  The condition $Dq^2=O(1)$ then shows
that $q^2$ has the same order as $\Omega_1$ and $\Omega_2$.  Above
$q\sim\sigL$, the exact kernels in Eqs.~\eqref{eq:Fph_partial_main} and
\eqref{eq:Fph_FC} behave as $\sigL/q$, so their momentum integrals grow
logarithmically with the two large frequencies.  The phonon and Coulomb
logarithms enter with opposite signs.  When their sum is attractive, its
coefficient is proportional to $g$, whereas the trial energy grows linearly with
the frequencies.  \apprev{The derivative of the trial term with respect to a
large frequency is therefore of order unity, while that of the attractive
logarithm has magnitude of order $g/\Omega_i$.  At a stationary point these
contributions balance, giving frequencies proportional to $g$.}}

\rev{We therefore introduce the positive dimensionless variational parameters
$x$ and $y$ through}
\begin{align}
\Omega_1&=g\,[x+o(1)],
&
\Omega_2&=g\,[y+o(1)],
\nonumber\\
\Omega_3&=O(1),
&
\nu&=O(1).
\label{eq:compact_frequency_scaling}
\end{align}
\rev{\apprev{At fixed $s>0$, the terms in
Eqs.~\eqref{eq:D11}--\eqref{eq:D12} have different orders in $g$.}  Since
$\nu=O(1)$ and $\Omega_1=O(g)$, the diffusive contribution
$\nu^2s/(2\Omega_1^2)$ is $O(g^{-2})$.  Independently,
$\Omega_1s,\Omega_2s\to\infty$, so
$e^{-\Omega_1s},e^{-\Omega_2s}\to0$.  The corresponding terms reduce to
$1/(2\Omega_1)$ and $1/(2\Omega_2)$, while the terms containing the finite
frequency $\Omega_3$ are also $O(g^{-2})$.  \apprev{For both
$(i,j)=(1,1)$ and $(1,2)$, the leading terms are}}
\begin{align}
\apprev{D_{ij}(s)}
&=\frac{1}{2\Omega_1}+\frac{1}{2\Omega_2}+o(g^{-1})
\nonumber\\
&=\frac{x+y}{2gxy}+o(g^{-1}).
\label{eq:compact_D_scaling}
\end{align}
\rev{The fixed-$s$ expansion is not uniform at $s=0$: one has
$D_{11}(0)=0$ exactly, and $D_{12}(0)$ is not obtained by setting $s=0$ in
Eq.~\eqref{eq:compact_D_scaling}.  The exponentials remain relevant only in
the narrow interval $s=O(g^{-1})$, where $\Omega_{1,2}s=O(1)$.  Since
$\bar\Omega_q=O(1)$ at fixed $\sigzero$, this interval also has width
$O(g^{-1})$ in the $\tau$ integration.  It can modify only terms smaller than
the retained $g\ln g$ and $g$ contributions.  The direct Coulomb term has no
time integration and uses the equal-time width, which must therefore be
expanded directly from Eq.~\eqref{eq:D120_positive}.  Together with
Eq.~\eqref{eq:Etrial}, this gives}
\begin{align}
D_{12}(0)&=\frac{1}{gy}+o(g^{-1}),
&
\Etr&=\frac{g}{2}(x+y)+o(g).
\label{eq:compact_equal_time_trial_scaling}
\end{align}
\rev{For the independently minimized one-polaron functional, the
strong-coupling balance is between a trial term proportional to $v$ and a
logarithmic interaction term of order $-g\ln v$.  This balance implies
$v=O(g)$; we therefore introduce the finite positive dimensionless
variational parameter $z$ through $v=gz$.  The second trial frequency remains
$w=O(1)$.  At fixed $s>0$, the diffusive term in
$D_{\rm pol}$ is $O(g^{-2})$, whereas $e^{-vs}\to0$.
Equations~\eqref{eq:Dpol} and~\eqref{eq:Etr_pol} then give}
\begin{align}
D_{\rm pol}(s)&=\frac{1}{gz}+o(g^{-1}),
\nonumber\\
\mathcal E_{\rm tr}^{\rm pol}&=\frac{gz}{2}+o(g).
\label{eq:compact_pol_scaling}
\end{align}
\rev{\apprev{The short-time interval is subleading for the same reason as in
the bipolaron calculation.}  The Gaussian
factor selects characteristic momenta $q_{\rm char}=O(\sqrt g)$.  At fixed
$\sigL$ and $\sigzero$, $q_{\rm char}/\sigL\to\infty$ and hence
$\bar\Omega_q\to1$ in the momentum region responsible for the leading
logarithm.  Dispersion therefore does not change the leading $g\ln g$
coefficient, although it still affects subleading terms and finite-coupling
results.}

\subsection{Momentum integrals}

\apprev{The Coulomb and phonon energies contain the momentum integrals}
\begin{align}
J_{\rm C}(D)
&\equiv\int_0^\infty C(q)e^{-Dq^2}\dd q,
\label{eq:JC_definition_app}
\\
J_{\rm ph}(D)
&\equiv\int_0^\infty F(q)e^{-Dq^2}\dd q.
\label{eq:Jph_definition_app}
\end{align}
\rev{\apprev{Here $D$ denotes the correlator width in the corresponding energy term.}  The
relevant widths were obtained in Eqs.~\eqref{eq:compact_D_scaling}--
\eqref{eq:compact_pol_scaling} and are all $O(g^{-1})$.  Both integrals can be
expressed through the single auxiliary integral}
\begin{equation}
\begin{aligned}
\mathcal C(d)
&=\int_0^\infty\frac{e^{-dp^2}}{1+p}\dd p
\\
&=\frac{e^{-d}}{2}
\left[\pi\operatorname{erfi}(\sqrt d)-\operatorname{Ei}(d)\right],
\qquad d>0.
\end{aligned}
\label{eq:C_boundary_app}
\end{equation}
\rev{Here $\operatorname{erfi}$ and $\operatorname{Ei}$ are the imaginary
error function and the exponential integral.  With $q=\sigL p$ and the
partial-fraction form~\eqref{eq:Fph_partial_main}, the original momentum
integrals are}
\begin{align}
J_{\rm C}(D)
&=\sigL\mathcal C(D\sigL^2),
\nonumber\\
J_{\rm ph}(D)
&=\frac{\sigzero}{\sigzero-1}\sigL\left[
\mathcal C(D\sigL^2)
-\frac{1}{\sigzero}
\mathcal C\!\left(\frac{D\sigL^2}{\sigzero^2}\right)
\right].
\label{eq:J_C_ph_C_relation}
\end{align}
\rev{The small-$d$ expansion is}
\begin{equation}
\mathcal C(d)=-\frac{1}{2}\ln d-\frac{\gamma_{\rm E}}{2}+o(1),
\qquad d\to0^+.
\end{equation}
\rev{Here $\gamma_{\rm E}$ is the Euler--Mascheroni constant.  Consequently,
as $D\to0^+$ at fixed $\sigL$ and $\sigzero$,}
\begin{align}
J_{\rm C}(D)
&=\frac{\sigL}{2}\ln\frac{1}{D\sigL^2}
-\frac{\gamma_{\rm E}\sigL}{2}+\apprev{\sigL o(1)},
\label{eq:JC_asymptotic}
\\
\apprev{J_{\rm ph}(D)}
&\apprev{{}=\sigL\!\left[\frac12\ln\frac{1}{D\sigL^2}
-\frac{\gamma_{\rm E}}2
-\frac{\ln\sigzero}{\sigzero-1}+o(1)\right].}
\label{eq:Jph_asymptotic}
\end{align}
\rev{The leading logarithms are identical.  Their finite difference is}
\begin{equation}
J_{\rm C}(D)-J_{\rm ph}(D)
\longrightarrow\frac{\sigL\ln\sigzero}{\sigzero-1}>0.
\label{eq:J_difference}
\end{equation}
\rev{Equation~\eqref{eq:J_difference} compares the Coulomb and phonon kernels
at the same Gaussian width.  \apprev{In the bipolaron energy, these integrals
have different coefficients, and the direct interaction uses a different width.}}

\subsection{Polaron and bipolaron energies}

\rev{\apprev{Using the leading widths and trial energies in
Eqs.~\eqref{eq:compact_D_scaling}--\eqref{eq:compact_pol_scaling} leaves only
the time integral $\int_0^\infty e^{-\tau}\dd\tau=1$.}
Equations~\eqref{eq:Epol_2D} and~\eqref{eq:nl2d_long_bp} therefore reduce to}
\begin{align}
\varepsilon_{\rm pol}
&=\mathcal E_{\rm tr}^{\rm pol}
-\alphaeff J_{\rm ph}(D_{\rm pol})+o(g),
\nonumber\\
\varepsilon_{\rm bp}
&=\Etr
-2\alphaeff J_{\rm ph}(D_{11})
-2\alphaeff J_{\rm ph}(D_{12})
\nonumber\\
&\quad
+2\alphaeff\frac{\sigzero}{\sigzero-1}
J_{\rm C}\!\left(D_{12}(0)\right)
+o(g).
\label{eq:compact_energy_accounting}
\end{align}
\rev{The first line follows from the one-polaron functional
\eqref{eq:Epol_2D}: it contains the one-polaron trial energy and its phonon
contribution.  In the second and third lines, the terms occur in the same
order as in the bipolaron functional~\eqref{eq:nl2d_long_bp}: $\Etr$ is the
trial energy, the two negative terms are the same-particle and cross phonon
contributions, and the positive term is the direct Coulomb repulsion.
Equation~\eqref{eq:compact_D_scaling} makes the two phonon widths equal only
at the leading strong-coupling order, while the Coulomb term uses the
distinct equal-time width $D_{12}(0)$.}

\apprev{Substitution of Eq.~\eqref{eq:Jph_asymptotic} into the one-polaron
energy gives}
\begin{equation}
\frac{\varepsilon_{\rm pol}}{g}
=\frac{z}{2}-\frac{1}{2}\ln\!\left(\frac{gz}{\sigL^2}\right)
+\frac{\gamma_{\rm E}}{2}
+\frac{\ln\sigzero}{\sigzero-1}+o(1).
\label{eq:Epol_asymptotic_z}
\end{equation}
\rev{Differentiation with respect to $z$ gives $z_*=1$, so the separately
optimized two-polaron reference is}
\begin{equation}
\frac{2\varepsilon_{\rm pol}}{g}
=-\ln\!\left(\frac{\alphaeff}{\sigL}\right)
+1+\gamma_{\rm E}
+\frac{2\ln\sigzero}{\sigzero-1}+o(1).
\label{eq:two_Epol_asymptotic}
\end{equation}
\apprev{For the bipolaron, using both expansions
\eqref{eq:JC_asymptotic}--\eqref{eq:Jph_asymptotic} gives}
\begin{align}
\frac{\varepsilon_{\rm bp}}{g}
&=-\frac{\sigzero-2}{\sigzero-1}
\ln\!\left(\frac{\alphaeff}{\sigL}\right)
+\Phi(x,y;\sigzero)
\nonumber\\
&\quad +\frac{\sigzero-2}{\sigzero-1}\gamma_{\rm E}
+\frac{4\ln\sigzero}{\sigzero-1}+o(1),
\label{eq:Ebp_asymptotic_xy}
\\
\Phi(x,y;\sigzero)
&=\frac{x+y}{2}+2\ln\frac{x+y}{2xy}
+\frac{\sigzero}{\sigzero-1}\ln y.
\label{eq:Phi_xy}
\end{align}
Solving the stationarity conditions
$\partial_x\Phi=\partial_y\Phi=0$ gives the positive solution
\begin{equation}
x_*=\frac{4(\sigzero-2)}{3\sigzero-4},
\qquad
y_*=\frac{2(\sigzero-2)^2}
{(\sigzero-1)(3\sigzero-4)}.
\label{eq:xy_stationary}
\end{equation}
This solution exists only for $\sigzero>2$.  At the stationary point,
\begin{align}
\apprev{\Phi_*(\sigzero)}
&\apprev{{}=\frac{\sigzero-2}{\sigzero-1}
+\frac{3\sigzero-4}{\sigzero-1}
\ln\!\frac{3\sigzero-4}{\sigzero-1}}
\nonumber\\
&\apprev{\quad-\frac{2(\sigzero-2)}{\sigzero-1}
\ln\!\frac{\sigzero-2}{\sigzero-1}
+\frac{6-5\sigzero}{\sigzero-1}\ln2.}
\label{eq:Phi_stationary}
\end{align}
Finally, subtracting the bipolaron energy from the two-polaron reference
gives
\begin{align}
\frac{\varepsilon_{\rm bind}}{g}
&=-\frac{1}{\sigzero-1}
\ln\!\left(\frac{\alphaeff}{\sigL}\right)
+\mathcal B(\sigzero)+o(1),
\label{eq:Ebind_asymptotic_final}
\\
\mathcal B(\sigzero)
&=1+\frac{\gamma_{\rm E}}{\sigzero-1}
-\frac{2\ln\sigzero}{\sigzero-1}
-\Phi_*(\sigzero).
\label{eq:B_sigma0_nl2d}
\end{align}
For $1<\sigzero\le2$, no positive stationary
solution~\eqref{eq:xy_stationary} exists.  For every finite
$\sigzero>2$, the coefficient $-1/(\sigzero-1)$ is negative.
Consequently the logarithm in Eq.~\eqref{eq:Ebind_asymptotic_final} eventually
drives the binding indicator below zero as $\alphaeff$ increases.  Setting the
leading expression to zero gives the formal upper-boundary estimate
\begin{equation}
\alpha_{\mathrm{eff}}^{+}
\sim\sigL\exp\!\left[(\sigzero-1)\mathcal B(\sigzero)\right].
\label{eq:alpha_upper_asymptotic}
\end{equation}
In contrast, if $\sigzero\to\infty$ is taken before the coupling is made
large, the coefficient of the negative logarithm vanishes and
$\mathcal B(\sigzero)\to\ln(32/27)>0$.  Thus every finite $\sigzero$ ultimately loses
binding at sufficiently large coupling, whereas the idealized
$\sigzero=\infty$ model remains bound in this asymptotic solution.

\section{Large-\texorpdfstring{$\sigL$}{sigma-L} reduction at the boundary meeting}
\label{app:nl2d_boundary_meeting}

\rev{This appendix derives the limiting point at which the lower and upper
binding boundaries meet.  At such a point, the separately optimized binding
indicator has a double zero as a function of the coupling:}
\begin{equation}
\rev{\varepsilon_{\rm bind}=0,
\qquad
\partial_{\alphaeff}\varepsilon_{\rm bind}=0.}
\label{eq:boundary_double_zero_app}
\end{equation}
\rev{The limiting procedure differs from those in
Appendices~\ref{app:zero_corner} and~\ref{app:nl2d_asymptotic}.
Appendix~\ref{app:zero_corner} takes $\sigzero\to\infty$ before
$\sigL\to\infty$ to obtain the lower binding boundary, whereas
Appendix~\ref{app:nl2d_asymptotic} takes $\alphaeff\to\infty$ at fixed
$\sigL$ to obtain the high-coupling upper boundary.  Here $\sigzero$ remains
finite while $\sigL\to\infty$, and the required scaling of $\alphaeff$ is
determined below.}

\subsection{Scaling at the boundary meeting}

\rev{Set $q=\sigL p$, so that $\dd q=\sigL\dd p$ and
$q/\sigL=p$.  The exact form factors and dispersion therefore retain their
full dependence on the finite scaled momentum $p$.  The interaction measure
becomes}
\begin{equation}
\rev{\alphaeff\dd q=(\alphaeff\sigL)\dd p=g\dd p,}
\label{eq:boundary_interaction_measure_app}
\end{equation}
\rev{\apprev{where $g$ is defined in Eq.~\eqref{eq:g_nl2d}.}}

\apprev{At $q=\sigL p$, finite Gaussian exponents require widths of order
$\sigL^{-2}$.  Within the hierarchy of two large bipolaron frequencies and
one large polaron frequency, Eqs.~\eqref{eq:compact_D_scaling}--\eqref{eq:compact_pol_scaling}
then give $v,\Omega_1,\Omega_2$ and the trial energies of order $\sigL^2$.
The remaining frequencies $w,\Omega_3,\nu$ stay finite.  The interaction
integrals retain a finite dependence on the scaled widths and have the
prefactor $g$ in Eq.~\eqref{eq:boundary_interaction_measure_app}.  Balancing
the trial and interaction terms therefore requires $g$ of order $\sigL^2$.
Equivalently, the widths are of order $g^{-1}$ and}

\begin{equation}
\apprev{q^2D=O\!\left(\frac{\sigL^2}{\alphaeff\sigL}\right)
=O\!\left(\frac{\sigL}{\alphaeff}\right),\qquad p=O(1).}
\label{eq:boundary_scaling_balance_app}
\end{equation}
\rev{A finite nonzero exponent in the $q=O(\sigL)$ region requires
$\alphaeff/\sigL=O(1)$.  We therefore set}
\begin{align}
\sigL&\to\infty,
&
\alphaeff&=\bar\alpha\sigL,
\nonumber\\
q&=\sigL p,
&
g&=\alphaeff\sigL=\bar\alpha\sigL^2.
\label{eq:boundary_double_scaling_app}
\end{align}
\rev{Consistently with the main-text analysis, we set $\rho=0$ in the reduced
functional, so $J_0(q\rho)=1$.  We use the finite scaled variational parameters
$x,y,z>0$ introduced in Eqs.~\eqref{eq:compact_frequency_scaling} and
\eqref{eq:compact_pol_scaling}.  \apprev{Here $\bar\alpha=\alphaeff/\sigL$
is held finite and positive.  The parameters $x,y,z$ must be optimized anew
with the full kernels; their stationary values in the fixed-$\sigL$ limit
are not used.}  The fixed-time reduction of
Appendix~\ref{app:nl2d_asymptotic} makes the leading widths independent of
$\tau$, while the shrinking interval $\tau=O(g^{-1})$ contributes only beyond
leading order.  Hence $\int_0^\infty e^{-\tau}\dd\tau=1$.  The exact kernels
retain their dependence on $p=q/\sigL$, but the phonon dispersion no longer
appears in the leading reduced energies because the widths have lost their
time dependence.}

\subsection{Reduced functional}

\rev{The scaled Coulomb integral satisfies
$\mathcal C(d)=J_{\rm C}(D)/\sigL$ for $d=D\sigL^2$; the function itself and
this relation are given in Eqs.~\eqref{eq:C_boundary_app} and
\eqref{eq:J_C_ph_C_relation}.  In the same notation, define the scaled phonon
integral by}
\begin{align}
\mathcal P(d;\sigzero)
&\mathrel{\apprev{\equiv}}\int_0^\infty
\frac{\sigzero p}{(1+p)(1+\sigzero p)}e^{-dp^2}\dd p
\apprev{.}
\label{eq:P_boundary_app}
\end{align}
\apprev{Its closed form follows directly from
Eq.~\eqref{eq:J_C_ph_C_relation} by dividing $J_{\rm ph}(D)$ by $\sigL$
and substituting $D\sigL^2=d$.}
\rev{The finite Gaussian widths entering these scaled integrals are defined
directly from the original correlators:}
\begin{align}
d_{\rm pol}
&\equiv\lim_{\sigL\to\infty}\sigL^2D_{\rm pol}(s)
=\frac{1}{\bar\alpha z},
\qquad s>0,
\nonumber\\
d_{\rm ph}
&\equiv\lim_{\sigL\to\infty}\sigL^2D_{11}(s)
=\lim_{\sigL\to\infty}\sigL^2D_{12}(s)
\nonumber\\
&=\frac{x+y}{2\bar\alpha xy},
\qquad s>0,
\nonumber\\
d_{\rm C}
&\equiv\lim_{\sigL\to\infty}\sigL^2D_{12}(0)
=\frac{1}{\bar\alpha y}.
\label{eq:boundary_scaled_widths_app}
\end{align}
\rev{The} two leading energies are
\begin{align}
e_{\rm pol}
&\equiv\lim_{\sigL\to\infty}\frac{\varepsilon_{\rm pol}}{\bar\alpha\sigL^2}
=\frac{z}{2}-\mathcal P(d_{\rm pol};\sigzero),
\label{eq:reduced_pol_app}
\\
e_{\rm bp}
&\equiv\lim_{\sigL\to\infty}\frac{\varepsilon_{\rm bp}}{\bar\alpha\sigL^2}
\nonumber\\
&=\frac{x+y}{2}-4\mathcal P(d_{\rm ph};\sigzero)
+2\frac{\sigzero}{\sigzero-1}\mathcal C(d_{\rm C}).
\label{eq:reduced_bp_app}
\end{align}
\rev{In Eq.~\eqref{eq:reduced_pol_app}, $z/2$ is the one-polaron trial energy
and $-\mathcal P$ is its phonon contribution.  In
Eq.~\eqref{eq:reduced_bp_app}, $(x+y)/2$ is the bipolaron trial energy,
$-4\mathcal P$ is the sum of the same-particle and cross phonon terms, whose
widths coincide at this order, and the final positive term is the direct
Coulomb repulsion.}

\subsection{Meeting conditions and numerical result}

\rev{For each fixed $(\bar\alpha,\sigzero)$, the one-polaron energy is
minimized over $z>0$ and the bipolaron energy independently over $x,y>0$.
\apprev{Their optimized difference is
$2\min_{z>0}e_{\rm pol}-\min_{x,y>0}e_{\rm bp}$.}}
\rev{At the meeting point this quantity and its derivative with respect to
$\bar\alpha$ vanish.  \apprev{The original binding energy has the prefactor
$g=\bar\alpha\sigL^2$.  Its derivative contributes a term proportional to
the reduced binding energy, which vanishes at the crossing.  The remaining
term is proportional to the derivative of the reduced binding energy, so
these conditions are equivalent to
Eq.~\eqref{eq:boundary_double_zero_app} at leading order.}  Together with the
three stationarity conditions, this gives}
\begin{align}
\partial_z e_{\rm pol}&=0,
&
\partial_x e_{\rm bp}&=\partial_y e_{\rm bp}=0,
\nonumber\\
2e_{\rm pol}-e_{\rm bp}&=0,
&
2\partial_{\bar\alpha} e_{\rm pol}-\partial_{\bar\alpha} e_{\rm bp}&=0.
\label{eq:reduced_boundary_system_app}
\end{align}
\rev{The first three equations minimize the two energies.  The fourth sets
their optimized difference to zero, and the fifth makes the optimized binding
indicator tangent to zero.  These five equations determine the five unknowns
$(z,x,y,\bar\alpha,\sigzero)$.  The last equation has the displayed form by
the envelope theorem because derivatives of the optimized variational
parameters do not contribute at a stationary point.  \apprev{The numerical solution of}}
Eq.~\eqref{eq:reduced_boundary_system_app} \apprev{is}
\begin{align}
\sigstar&=89.113728\ldots,
&\bar\alpha^{\mathrm c}&=91.762596\ldots,
\nonumber\\
z_*&=0.850464\ldots,
&x_*&=1.130855\ldots,
\nonumber\\
y_*&=0.570072\ldots.
\label{eq:reduced_boundary_solution_app}
\end{align}
\rev{The one-polaron curvature is positive, both eigenvalues of the bipolaron
Hessian in $(x,y)$ are positive, and the second derivative of the optimized
binding function with respect to $\bar\alpha$ is negative.  Thus the solution
is a local minimum of each reduced energy with respect to the retained
variables and a local maximum of the optimized reduced binding energy as a
function of $\bar\alpha$.  \apprev{The endpoint intervals obtained from
independent calculations at large finite $\sigL$ have the common overlap}}
\begin{equation}
\apprev{[89.1016,\,89.1162].}
\label{eq:boundary_finite_sigma_bracket_app}
\end{equation}
\apprev{The reduced result in Eq.~\eqref{eq:reduced_boundary_solution_app}
lies inside this overlap.  Corrections at finite $\sigL$ have not been
estimated separately.}

\section{Numerical threshold protocol}
\label{app:numerics}

\subsection{Trial-parameter reconstruction}

For the numerical optimization, define
\begin{equation}
\widetilde M=\frac{M}{m},
\quad
\widetilde K=\frac{K\ell_{\mathrm L}^2}{\hbar\omega_{\mathrm L}},
\quad
\widetilde\kappa=\frac{\kappa\ell_{\mathrm L}^2}{\hbar\omega_{\mathrm L}},
\quad
\widetilde\kappa'=\frac{\kappa'\ell_{\mathrm L}^2}{\hbar\omega_{\mathrm L}}.
\label{eq:trial_dimensionless_springs}
\end{equation}
The local parameters of Eq.~\eqref{eq:S_trial_original} are reconstructed from
the trial eigenfrequencies as
\begin{align}
\widetilde M
&=\frac{\Omega_1^2}{\nu^2}-1,
\nonumber\\
\widetilde K
&=\frac{\Omega_1^2-\Omega_2^2-\Omega_3^2}{4},
\nonumber\\
\widetilde\kappa+\widetilde\kappa'
&=\frac{\widetilde M\nu^2}{2},
\nonumber\\
(\widetilde\kappa-\widetilde\kappa')^2
&=\frac{\widetilde M}{16}
\Big[(\Omega_2^2-\Omega_3^2)^2
\nonumber\\
&\hspace{3.6em}-(\Omega_2^2+\Omega_3^2-2\nu^2)^2\Big].
\label{eq:trial_frequency_map}
\end{align}
The sign of $\widetilde\kappa-\widetilde\kappa'$ only interchanges the two
equivalent local spring assignments.  The stability conditions stated
below Eq.~\eqref{eq:S_trial_original} keep all reconstructed quantities real
and the complete quadratic form stable.

\subsection{Threshold search}

The lower and upper thresholds in Fig.~\ref{fig:monolayer-phase} were obtained
by minimizing the one-polaron energy \eqref{eq:Epol_2D} and the bipolaron
energy \eqref{eq:nl2d_long_bp} at a sequence of $\alphaeff$ values for each
fixed $(\sigzero,\sigL)$.  We monitored the dimensionless binding indicator
$\varepsilon_{\rm bind}=2\varepsilon_{\rm pol}-\varepsilon_{\rm bp}$ defined
below Eq.~\eqref{eq:Ebind_def}.  \apprev{In the zero searches,
$\varepsilon_{\rm bp}$ is the optimized energy of the localized stationary
solution, continued through the crossing with two polarons.  The binding
indicator can then have either sign.  Including the
dissociated state in the final energy comparison makes the variational
binding energy zero, rather than negative, outside the binding interval.}
The two thresholds are the zeros enclosing
the positive interval.  Each zero was first located between two calculated
points with opposite signs and then refined by bisection, with both energies
reoptimized at every coupling.  To reject sign changes caused by quadrature or
minimization error, \apprev{a point inside the binding interval} was accepted only when denser
quadrature and new optimization starts still gave
$\varepsilon_{\rm bind}>10^{-4}$.

\apprev{The endpoint searches used a bound-constrained limited-memory
quasi-Newton algorithm (L-BFGS-B) on logarithms of positive frequency
differences, with relative objective tolerance $10^{-12}$.
There were five one-polaron and six bipolaron starting configurations, plus
continuation from adjacent points.  The localized solution was separated
from dissociation by $\Omega_3\ge0.05$; solutions at this numerical boundary
were excluded from the endpoint search.  Composite Gauss--Legendre time
quadratures used 18 intervals on $0\le\tau\le30$, with 18--24 nodes per
interval, and 820--1200 momentum nodes on the rational mapping of
$[0,\infty)$ to $[0,1)$.  Endpoint checks increased these to 540 time and
1600 momentum nodes.  At $\sigL=50$ and $\sigzero=89.11621$, the positive
energy difference changed from $0.00586635$ to $0.00586619$ at fixed
optimized frequencies.  At $\sigL=1000$ and $\sigzero=89.125$, it changed
from $776.75$ to $1362.16$: the sign was stable.  The $10^{-4}$ criterion is
an acceptance threshold for the positive energy difference; the uncertainty
estimated from quadrature refinement can be larger.}

\apprev{For the universal functionals in Appendix~\ref{app:zero_corner},
the retarded integrals were evaluated in the variables $\tau^{1/3}$ and
$p\tau^{2/3}$, which regularize the small-time momentum integral.
The three quadrature levels used 6, 10, and 14 nodes per interval on nine
time intervals up to $\tau=30$, 60, 120, and 180 nodes in the transformed
momentum variable $p\tau^{2/3}$, and 80, 180, and 260 nodes
in the direct-interaction integral, respectively.  Localized solutions were
followed with $\Omega_3\ge0.03$ in scaled units; the threshold value
$\Omega_3\simeq0.814$ lies well inside this domain.  L-BFGS-B minimization
used relative objective tolerance $2\times10^{-13}$ and multiple initial
frequency sets.}

\rev{The \apprev{phase-diagram} calculations used $\rho=0$, with all four frequency
variables reoptimized at every parameter point.  As a check, representative
threshold, bound, and unbound points were recomputed both at fixed nonzero
$\rho$ and by jointly minimizing all five variational parameters.
\apprev{These checks included all 16 endpoint calculations with
$0.25\le\sigL\le1000$, separation profiles on
$10^{-7}\le\sigL\rho\le100$, and joint five-parameter minimizations.
Refined checks at $\sigL=0.25,1,5,20$ returned
$\sigL\rho<2\times10^{-5}$; the finest used 252 time and 820 momentum
nodes.  At $\sigL=5$ and $20$, setting the resulting small separation to
exactly zero at the same frequencies lowered the energy by
$3.43\times10^{-6}$ and $2.83\times10^{-6}$, respectively.}
\apprev{All localized threshold and bound solutions converged to zero
separation within numerical accuracy.  No lower-energy finite-separation
state was found within the searched trial class.}}

The point where the two boundaries meet was found by maximizing
$\varepsilon_{\rm bind}$ over
$\alphaeff$ at each $\sigzero$ and locating where this optimized maximum changes
sign.  Repeating the high-coupling calculation with denser time and momentum
quadratures reproduced Eq.~\eqref{eq:Ebind_asymptotic_final}.
\apprev{The endpoint intervals for $50\le\sigL\le1000$ are compared with the
reduced limiting value after Eq.~\eqref{eq:boundary_finite_sigma_bracket_app}.}

\bibliography{bipolaron_article_references}

\end{document}